\documentclass[journal,letterpaper]{IEEEtran}

\usepackage{amsmath} 
\usepackage{amssymb}                         
\usepackage[ruled,norelsize]{algorithm2e}    
\usepackage{bm}                              
\usepackage{color}
\usepackage[pdftex]{graphicx}                
\usepackage{nomencl}                         
\makenomenclature
\ifCLASSOPTIONcompsoc            
  \usepackage[caption=false,font=normalsize,labelfont=sf,textfont=sf]{subfig}
\else
  \usepackage[caption=false,font=footnotesize]{subfig}
\fi

\begin{document}

\title{\huge MetaRadar: Multi-target Detection for Reconfigurable Intelligent Surface Aided Radar Systems}

\author{Haobo~Zhang,~\IEEEmembership{Student Member,~IEEE,}
        Hongliang~Zhang,~\IEEEmembership{Member,~IEEE,}
        Boya~Di,~\IEEEmembership{Member,~IEEE,}
        Kaigui~Bian,~\IEEEmembership{Senior Member,~IEEE,}
        Zhu~Han,~\IEEEmembership{Fellow,~IEEE,}
        and~Lingyang~Song,~\IEEEmembership{Fellow,~IEEE}
        \thanks{Manuscript received March 22, 2021; revised September 2, 2021 and December 26, 2021; accepted Feburary 14, 2022. This work was supported in part by the National Key R\&D Project of China under Grant No. 2020YFB1807100, in part by the National Natural Science Foundation of China under Grants 61829101, 61941101, and 62032003, in part by Beijing Natural Science Foundation under Grant 4222005 and L212027, and in part by NSF CNS-2107216 and EARS-1839818. (Corresponding author: Lingyang Song.)}
        \thanks{H. Zhang, B. Di, and L. Song are with Department of Electronics, Peking University, Beijing 100871, China (e-mail: \{haobo.zhang,diboya,lingyang.song\}@pku.edu.cn).}%
        \thanks{H. Zhang is with Department of Electrical and Computer Engineering, Princeton University, NJ 08544, USA (e-mail: hz16@princeton.edu).}%
        \thanks{K. Bian is with Department of Computer Science, Peking University, Beijing, China (e-mail: bkg@pku.edu.cn).}%
        \thanks{Z. Han is with the Department of Electrical and Computer Engineering in the University of Houston, Houston, TX 77004 USA, and also with the Department of Computer Science and Engineering, Kyung Hee University, Seoul, South Korea, 446-701 (e-mail: zhan2@uh.edu).}
}

\maketitle

\begin{abstract}
  As a widely used localization and sensing technique, radars will play an important role in future wireless networks. However, the wireless channels between the radar and the targets are passively adopted by traditional radars, which limits the performance of target detection. To address this issue, we propose to use the reconfigurable intelligent surface~(RIS) to improve the detection accuracy of radar systems due to its capability to customize channel conditions by adjusting its phase shifts, which is referred to as MetaRadar. In such a system, it is challenging to jointly optimize both radar waveforms and RIS phase shifts in order to improve the multi-target detection performance. To tackle this challenge, we design a waveform and phase shift optimization (WPSO) algorithm to effectively solve the multi-target detection problem, and also analyze the performance of the proposed MetaRadar scheme theoretically. Simulation results show that the detection performance of the MetaRadar scheme is significantly better than that of the traditional radar schemes.
\end{abstract}


%
\IEEEpeerreviewmaketitle


\printnomenclature[2cm]

\section{Introduction}
\label{s_introduction}

Driven by the demand to support applications such as virtual reality~(VR) and autonomous driving, localization and sensing become crucial functions for future wireless systems~\cite{dang2020what,saad2020a}. To put this vision into practice, various sensing techniques which enable the perception of the surrounding environment are developed. Among various sensing techniques, radar technique which uses radio frequency~(RF) signals to detect and locate targets through reflected signals has attracted much attention due to many advantages such as environmental robustness, direct measurement of velocity, and low cost~\cite{campbell2018sensor}. Specifically, radars can directly measure the velocity of the target by leveraging the Doppler effect, which is especially useful for autonomous driving~\cite{jelena2018sensors}. In addition, with the development of RF CMOS and multiple-input-multiple-output~(MIMO) technology, radars are becoming more cost-efficient, which will lead to greater economic interests~\cite{igal2019the}.

In the literature, various works which optimize the waveforms of radar signals for detection performance enhancement have been proposed~\cite{jun2014target, feng2020joint}. Based on the optimization criterions, these works can be broadly classified into three types: detection probability, signal-to-interference-plus-noise ratio~(SINR), and relative entropy based schemes~\cite{long2019an}. The detection probability based schemes optimize the radar waveforms to directly maximize the detection probability given the constraint of false alarm probability, while the optimization problem is difficult to tackle due to the complicated relationship between the detection probability and the radar waveforms~\cite{gini2005waveform}. In the SINR based schemes, the SINRs of the echo signals from the targets are maximized because a large SINR usually leads to a high detection accuracy~\cite{yu2020mimo}. The relative entropy is another widely used criterion, which has been shown to be effective to evaluate the detection performance~\cite{kay2009waveform},~\cite{tang2009on}. In~\cite{tang2010mimo}, the authors considered the detection of a single target in colored noise, and derived the optimal radar waveforms that maximized the relative entropy. The multi-target detection scenario was investigated in~\cite{wang2018multi}. However, the propagation channels between the radar system and the targets are passively adopted in the aforementioned schemes, which limits the performance of the radar. Specifically, if the channel conditions are unfavorable, the radar signals will suffer heavy attenuation when propagating through the channels, resulting in a smaller SINR/relative entropy and a lower target detection accuracy. 

\textcolor{black}{To address this issue and improve the detection performance of the radar system, the RIS is a promising solution. The RIS is a type of planar metamaterial that can be used to effectively control the propagation channels~\cite{zhang2020reconfigurable, zhang2021beyond}. It is composed of a large number of elements with electrically tunable phase shifts, and thus the characteristics of reflection channels via the RIS elements can be changed by adjusting the phase shifts of these elements~\cite{renzo2019smart, hu2020reconfigurable}. By properly designing the phase shifts of RIS elements, the channel conditions between the radar antennas and the targets can be optimized to promote the detection performance of the radar systems\footnote{\textcolor{black}{The RIS-aided radar scheme can be utilized in the scenario where radar signals suffers heavy attenuation when propagating through the wireless channels. For example, the proposed scheme is especially suitable to extend the detection range of millimeter wave radars.}}.}

\begin{figure*}[!t]
  \centering
  \subfloat[]{
    \includegraphics[height=1.8in]{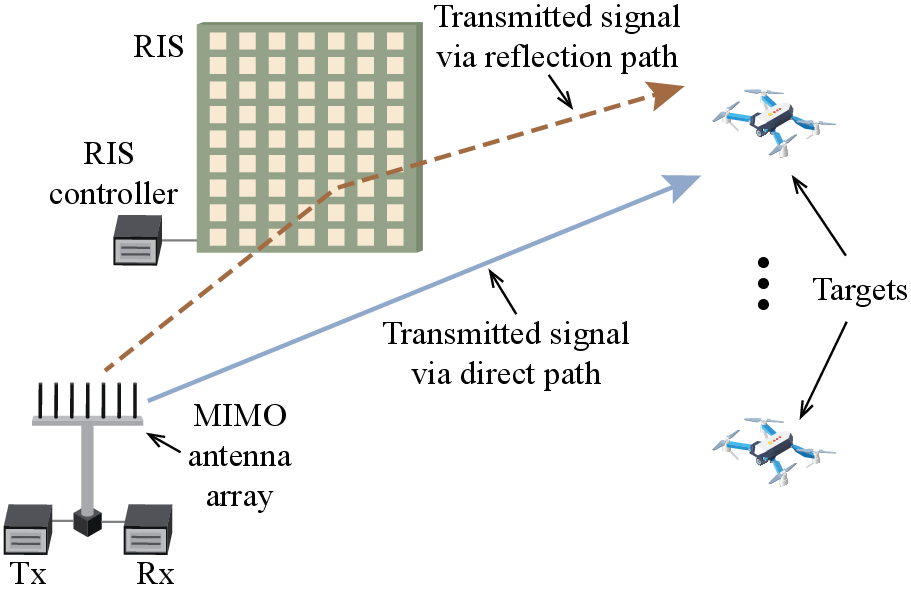}
  }
  \hspace{0.4in}
  \subfloat[]{
    \includegraphics[height=1.8in]{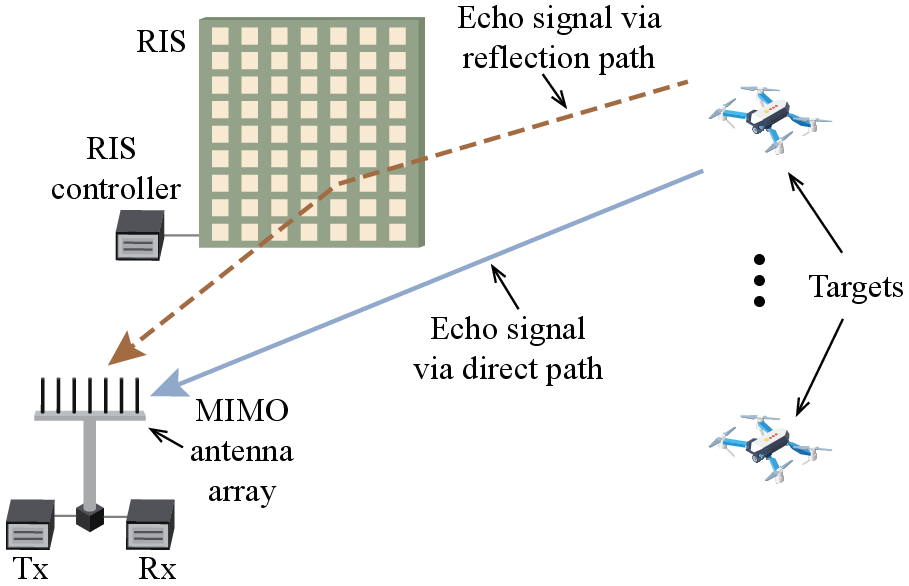}
  }
  \caption{A MetaRadar system: (a) Transmission mode; (b) Reception mode.}
  \label{f_model}
\end{figure*}

\textcolor{black}{Several RIS-aided radar schemes have been discussed in the literature. In~\cite{aubry2021reconfigurable}, the authors considered a scenario where there was no line-of-sight (N-LOS) link between the radar antennas and the target, and an RIS was deployed to enable the radar to detect the target in the NLOS areas. The authors in~\cite{buzzi2021radar} utilized an RIS to promote the received signal-to-noise ratio~(SNR) for enhanced detection capabilities of a single-antenna radar. The MIMO radar case was considered in~\cite{buzzi2021foundations} and \cite{lu2021intelligent}, where the phase shifts of the RIS were optimized to improve the radar performance. However, the aforementioned schemes all consider a single target scenario, which cannot be directly applied to the multi-target scenario. Besides, the radar waveforms are not optimized in these schemes, which results in limited improvement of radar performances. Different from these schemes, in this paper, we investigate the multi-target detection using an RIS-assisted radar, and jointly optimize the radar waveforms and RIS phase shifts in our proposed scheme.}

\textcolor{black}{Two challenges needs to be addressed in the proposed scheme. \emph{First}, to fully realize the benefits of the RIS in the multi-target detection scenario, the operations of the RIS need to be coordinated with the transceiver antennas, which complicates the system design. \emph{Second}, since radar waveforms and RIS phase shifts are jointly optimized for multi-target detection, the optimization problem is challenging because the two variables are coupled with each other. To handle these challenges, we propose the multi-target detection protocol for the MetaRadar, and formulate the multi-target detection problem which optimizes the radar waveform and RIS phase shifts based on the relative entropy criterion. The waveform and phase shift optimization~(WPSO) algorithm is design to efficiently tackle the formulated problem. In general, our contributions can be summarized as follows:}
\begin{itemize}
  \item We propose a multi-target detection protocol which coordinates the operations of the RIS and the radar antennas. The protocol runs in a cognitive manner, where the detection performance of the MetaRadar can be adaptively improved cycle by cycle.
  \item The multi-target detection problem is formulated using the criterion of relative entropy, where the radar waveforms and the RIS phase shifts are jointly optimized. The WPSO algorithm is designed to tackle the formulated multi-target detection problem, which can be used to efficiently derive the optimized radar waveforms and the RIS phase shifts.
  \item The convergence and complexity of the proposed WPSO algorithm are analyzed, and the relationship between the RIS phase shifts, radar configuration, and the detection performance is also discussed. We also verify the effectiveness of the proposed MetaRadar scheme through simulation.
  
\end{itemize}

The rest of this paper is organized as follows. In Section~\ref{s_model}, we describe the target detection scenario and introduce the model of the MetaRadar. The multi-target detection protocol is proposed in Section~\ref{s_protocol}. In Section~\ref{s_problem}, we formulate the waveform and RIS phase shift optimization problem. The WPSO algorithm is designed in Section~\ref{s_algorithm} to solve the formulated problem. In Section~\ref{s_analysis}, we provide the analysis of the proposed scheme. The simulation results are presented in Section~\ref{s_simulation}. Finally, we draw the conclusions in Section~\ref{s_conclusion}.

\section{System Model}
\label{s_model}

In this section, we first introduce the considered scenario in Subsection~\ref{ss_sd}, and then model the RIS, path gains, and the radar receiver in Subsections~\ref{ss_ris}, \ref{ss_pc}, and \ref{ss_rm}, respectively.

\subsection{Scenario Description}
\label{ss_sd}

We consider a multi-target detection scenario using the MetaRadar. As shown in Fig.~\ref{f_model}, the MetaRadar is composed of a transmitter~(Tx), a receiver~(Rx), a MIMO antenna array with $N$ antennas connected with the Tx and Rx, and an RIS with $M$ elements. By deploying an RIS in the radar system, we can create reflection paths between the MIMO antenna array and the targets, and thus the overall channel conditions between the array and targets can be improved by optimizing the phase shifts of the RIS.

\nomenclature{$N$}{Number of antennas in the MIMO array}%
\nomenclature{$M$}{Number of elements in the RIS}%

The MetaRadar functions in two modes, i.e., the transmission and reception modes. In the transmission mode, the Tx first generates signals according to designed waveforms, and then radiates the signals through the MIMO antenna array towards the targets via both direct and reflection paths, as illustrated in Fig.~\ref{f_model}~(a). Then, the MetaRadar converts to the reception mode, where the antenna array receives the echo signals reflected by the targets. The received signals will be delivered to the Rx in order to detect and locate targets.

\subsection{Reconfigurable Intelligent Surface}
\label{ss_ris}

\begin{figure}[!t]
  \centering
  \includegraphics[width=2.15in]{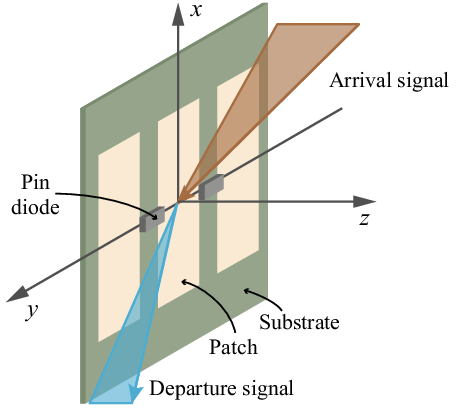}
  \caption{Reflection characteristics of an RIS element.}
  \label{f_element}
\end{figure}

The RIS is a type of planar material made up of many homogenous RIS elements. A programmable RIS element is illustrated in Fig.~\ref{f_element}. In an element, several metal patches are connected by the pin diodes and printed on the dielectric substrates. Each pin diode can be tuned to two states, i.e., \emph{ON} and \emph{OFF} states, leading to different states and reflection coefficients of the RIS element~\cite{di2020hybrid}. 

Suppose each RIS element has $N_s$ different reflection coefficients with same amplitude gain $\eta$ and $N_s$ different phase shifts which are uniformly distributed in the range $[0, 2\pi)$ with interval $\Delta s = \dfrac{2\pi}{N_s}$~\cite{di2020hybrid}. Therefore, the reflection coefficient of the $m$-th RIS element with phase shift $s_m$ can be expressed as
\begin{equation}
  r(s_m) = \eta e^{-j s}, s \in \left\{ i\Delta s | i = 1, \cdots, N_s\right\}.\label{def_r_m}
\end{equation}

\nomenclature{$N_s$}{Number of phase shifts of an RIS element}%
\nomenclature{$s_m$}{Phase shift of the $m$-th RIS element}%
\nomenclature{$r(s_m)$}{Reflection coefficient of the $m$-th RIS element with phase shift $s_m$}%

\subsection{Path Gains}
\label{ss_pc}

\begin{figure}[!t]
  \centering
  \includegraphics[width=3.5in]{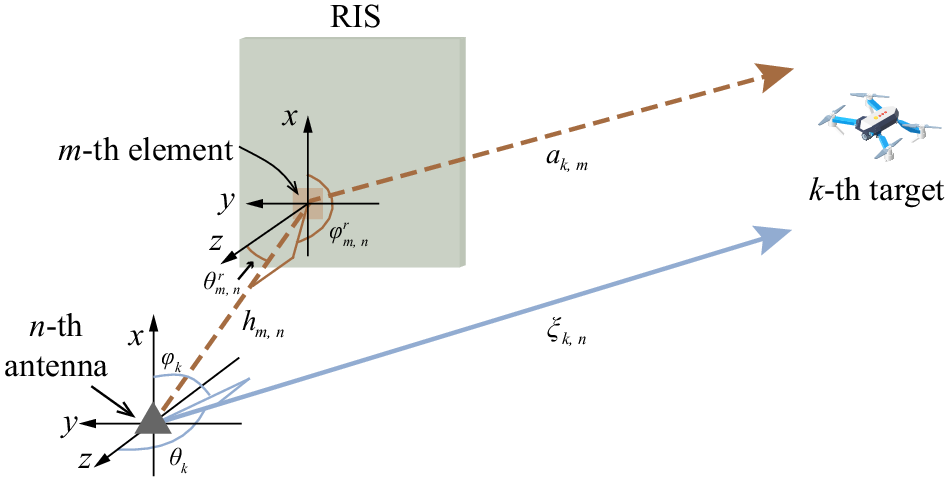}
  \textcolor{black}{\caption{Illustration of the main channel parameters.}}
  \label{f_channel_para}
\end{figure}

The reflection paths between the antenna array and the targets consists of two parts: the antenna-RIS path and the RIS-target path. More specifically, the path gain between the $n$-th antenna and the $m$-th RIS element is given by~\cite{elmossallamy2020reconfigurable, parsons1994the}
\begin{align}
  h_{m, n} =~&\dfrac{1}{\sqrt{4\pi}} \times \dfrac{\sqrt{G^A G^A_P(\theta^r_{m, n}, \varphi^r_{m, n}) G^R_P(\theta^r_{m, n}, \varphi^r_{m, n}) S^e} }{l_{m, n}} \notag\\
  &\times e^{- j 2 \pi l_{m, n} / \lambda},
\end{align}
where $G^A$ is the gain of an antenna. $(\theta^r_{m, n}, \varphi^r_{m, n})$ is the direction from the $m$-th element to the $n$-th antenna, as illustrated in Fig.~\ref{f_channel_para}. $G^A_P(\theta^r_{m, n}, \varphi^r_{m, n})$ is the normalized radiation pattern towards direction $(\theta^r_{m, n}, \varphi^r_{m, n})$. $G^R_P(\theta^r_{m, n}, \varphi^r_{m, n})$ is normalized radiation pattern of an RIS element towards direction $(\theta^r_{m, n}, \varphi^r_{m, n})$. $S^e$ is the area of an RIS element. $l_{m, n}$ is the distance between the $n$-th antenna and the $m$-th RIS element. $\lambda$ is the wavelength of the carrier signal. Based on~\cite{tang2020wireless}, the normalized radiation pattern $G^R_P(\theta, \varphi)$ can be modeled as
\begin{equation}
  G^R_P(\theta, \varphi) =
  \begin{cases}
    \text{cos}^3(\theta), &\theta \in [0, \pi/2], \varphi \in [0, 2\pi],\\
    0, &\theta \in (\pi/2, \pi], \varphi \in [0, 2\pi].
  \end{cases}
\end{equation}

\nomenclature{$h_{m, n}$}{Path gain between the $n$-th antenna and the $m$-th RIS element}%
\nomenclature{$G^A$}{Gain of an antenna}%
\nomenclature{$G^A_P(\theta, \varphi)$}{Normalized radiation pattern of an antenna towards direction $(\theta, \varphi)$}%
\nomenclature{$(\theta^r_{m, n}, \varphi^r_{m, n})$}{Direction from the $m$-th element to the $n$-th antenna}%
\nomenclature{$l_{m, n}$}{Distance between the $n$-th antenna and the $m$-th RIS element}%
\nomenclature{$\lambda$}{Wavelength of the carrier signal}%

Since targets are in the farfield of RIS\footnote{The distance between the targets and the MetaRadar is greater than $2D^2/\lambda$, where $D$ is the size of the RIS.}, signals reflected by the RIS can be viewed as a plane wave at the location of each target~\cite{skolnik2008radar}. Therefore, the amplitude gain and the relative phase delay of the $m$-th RIS element comparing with the first element towards the $k$-th target is given by
\begin{equation}
  a_{k, m} = \sqrt{G^R G^R_P(\theta_k, \varphi_k)} e^{j\bm{e}_k\bm{p}^e_m},
\end{equation}
where $G^R$ is the gain of an RIS element, $(\theta_k, \varphi_k)$ is the direction of the $k$-th target, $\bm{e}_k$ is the wave vector with wavelength $\lambda$ and direction towards the $k$-th target, and $\bm{p}^e_m$ is the relative position of the $m$-th RIS element compared with the first element. The vector $\bm{a}_k = (a_{k, 1}, \cdots, a_{k, M})$ is referred to as the steering vector of the RIS for the $k$-th target~\cite{skolnik2001Introduction}.

\nomenclature{$G^R$}{Gain of an RIS element}%
\nomenclature{$G^R_P(\theta, \varphi)$}{Normalized radiation pattern of an RIS element towards direction $(\theta, \varphi)$}%
\nomenclature{$(\theta_k, \varphi_k)$}{Direction towards the $k$-th target}%
\nomenclature{$\bm{a}_k$}{Steering vector of the RIS for the $k$-th target}%

Therefore, the gain of the reflection path from the antenna array to the directions of targets can be expressed as
\begin{equation}
  \bm{B}(\bm{s}) = \bm{A} \bm{R}(\bm{s}) \bm{H},\label{def_b}
\end{equation}
where $\bm{B}(\bm{s}) = (\bm{b}_1(\bm{s}), \cdots, \bm{b}_K(\bm{s}))^{\text{T}} \in \mathbb{C}^{K \times N}$, with $K$ being the number of targets. $\bm{A} = (\bm{a}_1, \cdots, \bm{a}_K)^{\text{T}} \in \mathbb{C}^{K \times M}$ is the steering vector of the RIS. $\bm{H} = [h_{m, n}] \in \mathbb{C}^{M \times N}$ is the matrix of channel gain between the antenna array and the RIS. $\bm{R}(\bm{s}) = \text{diag}(\bm{r}(\bm{s})) \in \mathbb{C}^{M \times M}$ is the reflection coefficient matrix of the RIS under phase shift vector $\bm{s} = (s_1, \cdots, s_M)$, and $\bm{r}(\bm{s}) = (r(s_1), \cdots, r(s_M))$ is the reflection coefficient vector of the RIS.

\nomenclature{$\bm{s}$}{Phase shift vector that contains the phase shifts of all the RIS elements}%
\nomenclature{$\bm{B}(\bm{s})$}{Gain of the reflection path from the antenna array to the directions of targets under phase shift vector $\bm{s}$}%
\nomenclature{$K$}{Number of targets}%
\nomenclature{$\bm{b}_k(\bm{s})$}{Gain of the reflection path from the antenna array to the direction of the $k$-th target under phase shift vector $\bm{s}$}%

Similarly, the gain of the reflection path from the targets to the antenna array under phase shift vector $\bm{s}$ can be expressed as
\begin{equation}
  \bm{B}^{\text{T}}(\bm{s}) = \bm{H}^{\text{T}} \bm{R}^{\text{T}}(\bm{s}) \bm{A}^{\text{T}}.
\end{equation}

\textcolor{black}{In addition, we assume the antenna array and the RIS are closely spaced, and the targets are in the farfield of both the antenna array and the RIS. In other words, the distance between the antenna array and the RIS is much smaller than that between the antenna array (or RIS) and the targets. Thus, the direction of a target is the same for both the RIS element and the MIMO element\footnote{\textcolor{black}{In practice, the farfield constraint $d_{n, k} > 2 D^2_{m, n}/\lambda$ is easy to be satisfied. For example, suppose the working frequency of the radar is $3$GHz, the size of the RIS is $1$m$\times 1$m ($20\times 20$ elements), the size of the MIMO array is $0.2$m $\times 0.2$m ($4\times 4$ elements), and the distance between the RIS center and the MIMO array center is $1$m ($10\lambda$). Thus, the maximum distance between a MIMO antenna and an RIS element is about $2$m, which indicates that when $d_{n, k} > 80$m, the direction of a target is the same for all the RIS elements and the MIMO antennas. Since the radars working at $3$GHz are typically used for surveillance with a few hundreds of meter detection range~\cite{liu2020jointradar},~\cite{reed2016on}, this constraint is satisfied in practice, indicating that it is reasonable to assume the direction of a target is the same for both the RIS element and the MIMO element.}}, and the gain of the direct path from the $n$-th antenna to the $k$-th target can be given by}
\begin{equation}
  \xi_{k, n} = \sqrt{G^A G^A_P(\theta_k, \varphi_k)} e^{j\bm{e}_k \bm{p}^a_n},\label{def_xi_k_n}
\end{equation}
\textcolor{black}{where $G^A_P(\theta_k, \phi_k)$ is the normalized radiation pattern of the antenna in the MIMO array towards direction $(\theta_k, \phi_k)$, $\bm{p}^a_n$ is the relative position of the $n$-th antenna compared with the first antenna. Consequently, the steering vector of the antenna array towards the direction of the $k$-th target is $\bm{\xi}_k = (\xi_{k, 1}, \cdots, \xi_{k, N})$, and the steering vector of the antenna array towards the directions of targets is $\bm{\Xi} = (\bm{\xi}_1, \cdots, \bm{\xi}_K)^{\text{T}}$.} 

\nomenclature{$\bm{\Xi}$}{Gain of the direct path from the antenna array to the targets}%

\subsection{Receiver Model}
\label{ss_rm}

\textcolor{black}{The functionality of the Rx is to detect targets using the received echo signals, and the multi-target detection is performed using the multiple hypotheses testing techniques. Specifically, we first form multiple hypothesis to represent different detection results. During the detection process, we update the probability of each hypothesis using the received signals. When the detection process ends, the hypothesis with the highest probability will be selected as the detection result.}

Before defining the hypotheses, we first introduce some assumptions on the number and locations of targets. \textcolor{black}{Specifically, we consider a practical scenario, where the directions and the number $K$ of targets are unknown, and the number $K$ is in the range $[0, K_M]$, where $K_M$ is a positive constant.} Besides, the location of the $k$-th target is represented by the direction $(\theta_k, \varphi_k)$ and range $d_k$. We discretize the space of interest into $I$ angular grids denoted by $\mathcal{I} \in \{1, \cdots, I\}$, and a target is located in one of the grids. Given the number of targets and the direction of each target, the range of each target can be estimated based on the received signals, which will be discussed in Section~\ref{s_protocol}. This means that a hypothesis only needs to contain the information of the number and directions of targets. Consequently, we introduce indicative vector $\bm{i}(U) = (i_1(U), \cdots, i_{N(U)}(U))$ to represent hypothesis $U$, which means that there are $N(U)$ targets, and the $k$-th target is in the $i_k(U)$-th grid.

\nomenclature{$d_k$}{Range of the $k$-th target}%
\nomenclature{$I$}{Number of angular grids}%
\nomenclature{$U$}{A hypothesis representing that there are $N(U)$ targets and the $k$-th target is in the $i_k(U)$-th grid}%

The prior probabilities of hypotheses have to be initialed before the detection process. Let $p^1(U)$ denote the prior probability of hypothesis $U$.
Since $K$ is unknown and $K$ is the range $[0, K_M]$, without loss of generality, we assume that $K$ follows uniform distribution in range $[0, 1, \cdots, K_M]$, which is a commonly used assumption when there is no prior knowledge of the number of targets~\cite{wang2018multi}. Thus, the prior probability of hypothesis $U$ can be given by
\begin{equation}
  p^1(U) = \dfrac{1}{K_M + 1}\times \dfrac{1}{J(U)},
\end{equation}
where $J(U)$ denotes the number of hypotheses with $N(U)$ targets.

\nomenclature{$p^c(U_j)$}{Prior probability of hypothesis $U_j$ in the $c$-th cycle}%
\nomenclature{$p^{(c)}(\bm{y}^{(c)}|U_j)$}{Probability to receive $\bm{y}^{(c)}$ given hypothesis $U_j$}%
\nomenclature{$|| \bm{\beta} ||$}{Euclidean norm of vector $\bm{\beta}$}%

Based on Bayes' theorem, the prior probabilities of hypotheses can be updated by exploiting the information in the received signals, and the decisions of target detection can be made using these prior probabilities. Details of the probability updating process will be introduced in Section~\ref{s_protocol}.

\section{Multi-target Detection Protocol}
\label{s_protocol}

In this section, we propose a multi-target detection protocol to coordinate the operations of Tx, Rx, antenna array, and RIS in the detection process. We divide the timeline in the detection process into cycles with duration $\delta_C$, and the probabilities of all the hypotheses will be updated cycle by cycle. After $C$ cycles, the detection process will terminate, and the hypothesis with the highest probability will be chosen as the correct hypothesis~\cite{levy2008principles}.

\begin{figure*}[!t]
  \centering
  \includegraphics[width=5.5in]{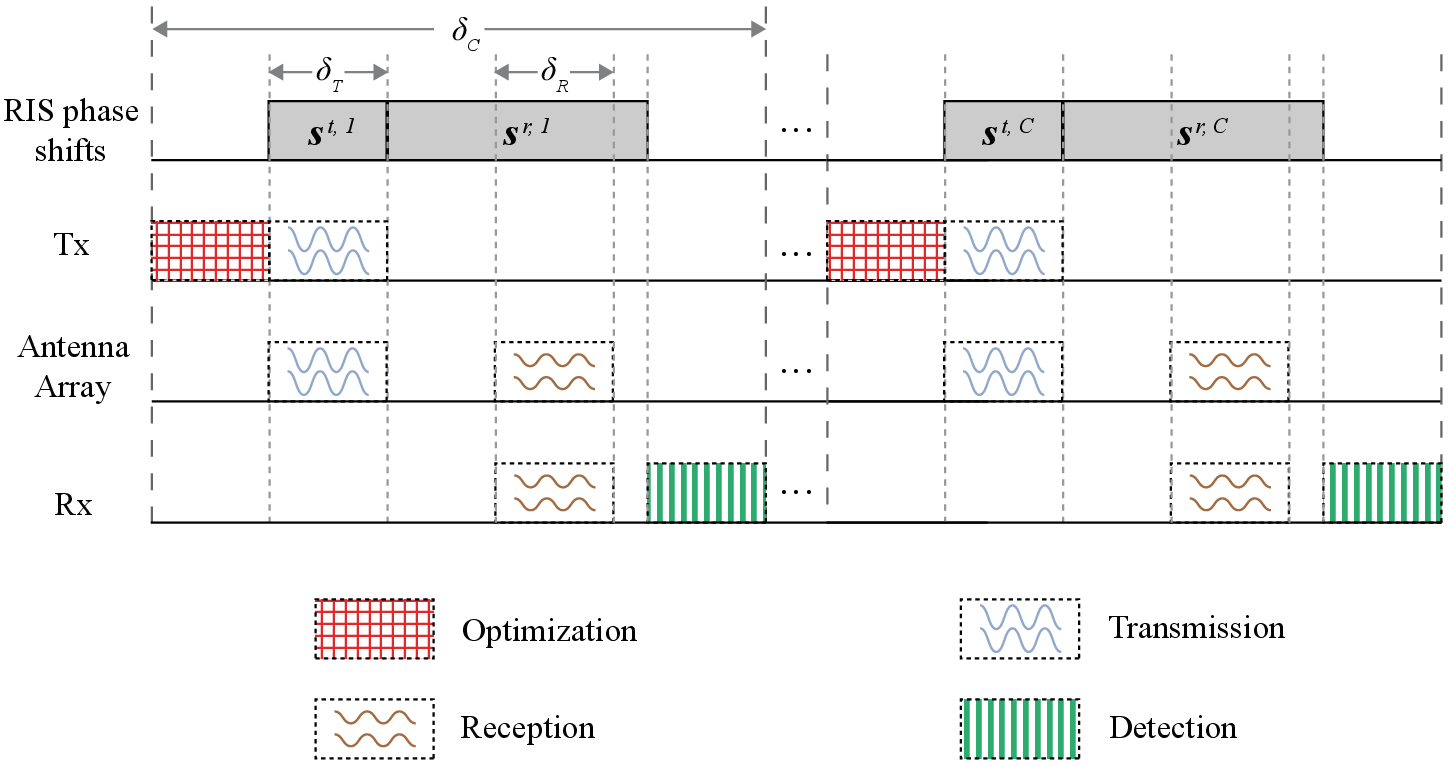}
  \caption{Multi-target detection protocol.}
  \label{f_protocol}
\end{figure*}

As illustrated in Fig.~\ref{f_protocol}, each cycle contains four steps, i.e., the optimization, transmission, reception, and the detection steps.

\emph{1) Optimization:} In this step, radar waveforms and the RIS phase shifts are optimized by the Tx based on the signals received in previous cycles. Let $\bm{W}^c \in \mathbb{C}^{N \times L}$ denote the generated waveform matrix in the $c$-th cycle, with $L$ being the number of snapshots. Besides, the optimized vectors of RIS phase shifts in the transmission and reception steps in the $c$-th cycle are denoted by $\bm{s}^{t, c} = (s^{t, c}_1, \cdots, s^{t, c}_M)$ and $\bm{s}^{r, c} = (s^{r, c}_1, \cdots, s^{r, c}_M)$, respectively\footnote{\textcolor{black}{The RIS phase shifts $\bm{s}^{t, c}$ and $\bm{s}^{r, c}$ are sent to the RIS controller in the optimization step in order to avoid the signaling cost in the transmission and reception steps.}}. Details of the optimization problem will be introduced in Section~\ref{s_problem}. As for the first cycle, the waveform matrix $\bm{W}^1$ and RIS phase shift vectors $\bm{s}^{t, 1}$ and $\bm{s}^{r, 1}$ are randomly generated.

\nomenclature{$\bm{W}^c$}{Radar waveform matrix in the $c$-th cycle}%
\nomenclature{$L$}{Number of snapshots of the radar waveform}%
\nomenclature{$\bm{s}^{t, c}$}{RIS phase shift vector in the transmission step in the $c$-th cycle}%

\emph{2) Transmission:} The generated radar waveforms $\bm{W}^c$ are emitted by the antenna array towards the RIS in this step. The RIS phase shifts are set as $\bm{s}^{t, c}$ during the whole step\footnote{\textcolor{black}{The duration of the adjustment of RIS phases is much smaller than that of the transmission step $\delta_T$~\cite{zhang2021a, tang2020mimo}. For example, when the target is $150$m away from the radar, the duration $\delta_T$ is $1\mu$s. In contrast, the phases of the RIS can be changed within $12.5$ns~\cite{zhang2021a}, which is much smaller than $\delta_T$. The price to adjust the phases is also acceptable because the control circuit of the RIS is very simple, and the main cost of the circuit is the low-cost FPGA module.}}. Based on (\ref{def_b}) and~(\ref{def_xi_k_n}), the transmitted signals towards the directions of the targets can be expressed as
\begin{equation}
  \bm{Y}^{t, c} = (\bm{B}(\bm{s}^{t, c}) + \bm{\Xi})\bm{W}^c,
\end{equation}
where $\bm{Y}^{t, c} = (\bm{y}^{t, c}_1, \cdots, \bm{y}^{t, c}_K)^{\text{T}} \in \mathbb{C}^{K\times L}$, with $\bm{y}^{t, c}_k$ being the transmitted signal towards the direction of the $k$-th target.

\nomenclature{$\bm{Y}^{t, c}$}{Transmitted signal towards the directions of the targets in the $c$-th cycle}%
\nomenclature{$\bm{y}^{t, c}_k$}{Transmitted signal towards the direction of the $k$-th target in the $c$-th cycle}%

\emph{3) Reception:} After the waveforms $\bm{W}^c$ are transmitted, the phase shifts of the RIS are set as $\bm{s}^{r, c}$, and the antenna array listens for the echo signals from the targets. Since the distances between the targets and the radar can be different, the echo signals from different targets may have different delays. Suppose that the echo signals received by the antenna array from the the $k$-th target delays $l_k$ snapshots compared with the transmitted signals. The delayed signal from the $k$-th target is given by
\begin{equation}
  \bm{y}^{d, c}_k = \bm{y}^{t, c}_k \bm{J}_k,
\end{equation}
where $\bm{J}_k$ is a shift matrix with size $L \times L_R$, with $L_R$ being the number of snapshots of the received signal. The $(l \times l')$-th element of $\bm{J}_k$ satisfies
\begin{equation}
  \bm{J}_k(l, l') =\begin{cases}
    1, l'-l-l_k + L^m = 0,\\
    0, \text{otherwise},
  \end{cases}
\end{equation}
where $L^m$ is the minimum delay. Besides, we assume that $L^m > L$ so that the received signals will not overlap with the transmitted signals. Consequently, the signals received by the antenna array can be expressed as\footnote{\textcolor{black}{The gain of the double cascaded channel $\bm{b}^{\text{T}}_k(\bm{s}^{r, c})\bm{b}_k(\bm{s}^{t, c})$ can be comparable with that of the LOS channel $\bm{\xi}^{\text{T}}_k \bm{\xi}_k$ from the antenna array to the antenna array via the targets. According to~\cite{tang2020wireless}, the gain of the double cascaded channel can be roughly expressed as $A/(d^2_1 d^2_2)$, where $A$ is a parameter related to the gain of the RIS, $d_1$ denotes the distance between the antenna and the RIS, and $d_2$ denotes the distance between the RIS and the targets. Besides, the gain of the LOS channel can be approximated by $B/d^2_3$, where $B$ is a parameter related to the gain of the antenna, and $d_3$ denote the distance between the antenna and the targets. Since we assume that the distance between the antenna and the RIS is much smaller than the distance between the RIS/antenna array and the target, we have $d_1 \ll d_2 \approx d_3$. In addition, as we assume that the antenna array are close to the RIS, by carefully designing the phase shifts of the RIS, the gain of the RIS can be promoted to satisfy $A/d^2_1 \ge B$. Thus, we can assure that the gain of the double cascaded channel is the same or even larger than that of the LOS channel.}}
\begin{align}
  \bm{Y}^c = \sum^K_{k=1} \gamma_k \left( \bm{b}^{\text{T}}_k(\bm{s}^{r, c}) + \bm{\xi}^{\text{T}}_k \right) \left(\bm{b}_k(\bm{s}^{t, c}) +  \bm{\xi}_k\right)\bm{W}^c \bm{J}_k + \bm{V}^{c},\label{def_Y_c}
\end{align}
where $\gamma_k$ is the response of the $k$-th target characterizing the reflection and channel propagation effects related to the $k$-th target~\cite{yu2020mimo}, and $\bm{V}^c \in \mathbb{C}^{L \times L_R}$ denotes the residual term which includes noise and interferences from the environment~\cite{zhang2021reconfigurable, li2008range}. For simplicity, we assume that the rows of $\bm{V}^{c}$ follow independent and identically distributed circularly symmetric complex Gaussian distribution with mean zero and covariance matrix $\sigma^2 \bm{I}^{L_R \times L_R}$~\cite{wang2018multi}. Let $\bm{y}^c = \text{vec}(\bm{Y}^c)$ denote the received signal vector, which is given by
\begin{equation}
  \bm{y}^c = \bm{F}^c \bm{\gamma} + \bm{v}^c,\label{def_y_c}
\end{equation}
where the response vector $\bm{\gamma} = (\gamma_1, \cdots, \gamma_K)^{\text{T}} \in \mathbb{C}^{K\times 1}$, the residual vector $\bm{v}^c = \text{vec}(\bm{V}^c) \in \mathbb{C}^{NL_R\times 1}$, and $\bm{F}^c = (\bm{Q}^c_1(\bm{s}^{t, c}, \bm{s}^{r, c}) \bm{w}^c, \cdots, \bm{Q}^c_K(\bm{s}^{t, c}, \bm{s}^{r, c}) \bm{w}^c) \in \mathbb{C}^{NL_R\times K}$ with $\bm{Q}^c_k(\bm{s}^{t, c}, \bm{s}^{r, c}) = \bm{J}^{\text{T}}_k\otimes ((\bm{b}^{\text{T}}_k(\bm{s}^{r, c})+\bm{\xi}^{\text{T}}_k) (\bm{b}_k(\bm{s}^{t, c}) + \bm{\xi}_k)) \in \mathbb{C}^{NL_R\times NL}$ and $\bm{w}^c = \text{vec}(\bm{W}^c) \in \mathbb{C}^{NL\times 1}$.

\nomenclature{$\bm{s}^{r, c}$}{RIS phase shift vector in the reception step in the $c$-th cycle}%
\nomenclature{$l_k$}{Delay of the echo signals received by the antenna array from the $k$-th target compared with the transmitted signals}%
\nomenclature{$\bm{y}^{d, c}_k$}{Delayed signal from the $k$-th target in the $c$-th cycle}%
\nomenclature{$L_R$}{Number of snapshots of the received signal}%
\nomenclature{$\bm{J}_k$}{Shift matrix of the $k$-th target which shifts signals from $\bm{y}^{t, c}_k$ to $\bm{y}^{d, c}_k$}%
\nomenclature{$\bm{Y}^c$}{Signal received by the antenna array in the $c$-th cycle}%
\nomenclature{$\gamma_k$}{Response of the $k$-th target characterizing the reflection and channel propagation effects related to the $k$-th target}%
\nomenclature{$\bm{V}^c$}{Residual term in the $c$-th cycle}%
\nomenclature{$\sigma^2$}{Variance of the residual term}%
\nomenclature{$\bm{y}^c$}{Vectorization of the matrix $\bm{Y}^c$}%
\nomenclature{$\bm{\gamma}$}{Vector of responses of all the targets}%
\nomenclature{$\bm{v}^c$}{Vectorization of the residual term $\bm{V}^c$}%
\nomenclature{$\bm{w}^c$}{Vectorization of the radar waveform $\bm{W}^c$}%

\emph{4) Detection:} The received signal $\bm{y}^c$ will then be processed by the Rx to update the probabilities of hypotheses for multi-target detection. Suppose the number of targets is $K$, we can form $J_K$ hypotheses, and we have
\begin{equation}
  j_{K, I} = \binom{I+K-1}{K}.
\end{equation}
Since $0 \le K \le K_M$, there are $J = J_0 + J_1 + \cdots + J_{K_M}$ hypotheses, denoted by $\{U_0, U_1, \cdots, U_{J-1}\}$. Given signal $\bm{y}^{(c)} = (\bm{y}^1, \cdots, \bm{y}^{c})^{\text{T}}$ received in previous $c$ cycles, the prior probability of hypothesis $U_j$ in the $(c+1)$-th cycle can be expressed as~\cite{zhang2020metaradar}
\begin{equation}
  p^{c+1}(U_j) = \dfrac{p^1(U_j)p^{(c)}(\bm{y}^{(c)}|U_j)}{\sum^J_{j=0} p^1(U_j)p^{(c)}(\bm{y}^{(c)}|U_j)}.\label{p_c}
\end{equation}
where $p^{(c)}(\bm{y}^{(c)}|U_j)$ denote the probability to receive $\bm{y}^{(c)}$ given hypothesis $U_j$, which can be given by
\begin{align}
  &p^{(c)}(\bm{y}^{(c)}|U_j)\notag\\
  =& \prod^{c}_{i=1} p^i(\bm{y}^i|U_j) \notag\\
  =& \prod^{c}_{i=1}\! \dfrac{1}{(\pi \sigma^2)^{NL_R}} \exp\!\left(\!{-\dfrac{||\bm{y}^i\! -\! \overline{\bm{y}}^i(U_j,\! \{\hat{l}_k\}^{(c), j},\! \hat{\bm{\gamma}}^{(c), j})||^2}{\sigma^2}}\!\right),\label{def_p_y_c}
\end{align}
where $\{\hat{l}_k\}^{(c), j}$ and $\hat{\bm{\gamma}}^{(c), j}$ denote the estimated delays and responses given hypothesis $U_j$, respectively. $\overline{\bm{y}}^i(U_j, \{\hat{l}_k\}^{(c), j}, \hat{\bm{\gamma}}^{(c), j})$ denotes the mean signals received by the antenna array under hypothesis $U_j$, delays $\{\hat{l}_k\}^{(c), j}$, and responses $\hat{\bm{\gamma}}^{(c), j}$ in the $i$-th cycle. \textcolor{black}{Note that $U_0$ denotes the hypothesis when $K = 0$, and thus $\overline{\bm{y}}^i(U_j, \{\hat{l}_k\}^{(c), j}, \hat{\bm{\gamma}}^{(c), j}) = \bm{0}, \forall i$. When $j\ne 0$, delays $\{\hat{l}_k\}^{(c), j}$ and responses $\hat{\bm{\gamma}}^{(c), j}$ can be estimated jointly using the maximum likelihood estimation method. Specifically, given delays $\{l_k\}$, the maximum likelihood estimation of responses $\tilde{\bm{\gamma}}^{(c), j}(\{l_k\})$ can be expressed as}
\textcolor{black}{\begin{equation}
  \tilde{\bm{\gamma}}^{(c), j}(\{l_k\}) = \left((\bm{F}^{(c)})^{\text{H}}\bm{F}^{(c)}\right)^{-1} (\bm{F}^{(c)})^{\text{H}}\bm{y}^{(c)},\label{e_gamma_tilde2}
\end{equation}}
\textcolor{black}{where $\bm{F}^{(c)} = (\bm{F}^{1}, \cdots, \bm{F}^{c})^{\text{T}}$ is calculated using delays $\{l_k\}$. Next, based on~(\ref{e_gamma_tilde2}), the maximum likelihood estimation of the delays can be given by}
\begin{align}
  &\{\hat{l}_k\}^{(c), j} \notag\\
  =& \arg\max_{\{l_k\}} p^{(c)}\left(\bm{y}^{(c)}|U_j, \{l_k\}\right)\notag\\
  =& \prod^{c}_{i=1} \dfrac{1}{(\pi \sigma^2)^{NL_R}} \exp\left({-\dfrac{||\bm{y}^i - \overline{\bm{y}}^i(U_j, \{l_k\}, \hat{\bm{\gamma}}^{(c), j})||^2}{\sigma^2}}\right),
\end{align}
where $p^{(c)}\left(\bm{y}^{(c)}|U_j, \{l_k\}\right)$ denotes the probability to receive $\bm{y}^{(c)}$ given hypothesis $U_j$, delays $\{l_k\}$, and responses $\tilde{\bm{\gamma}}^{(c), j}(\{l_k\})$. 
Finally, using the estimated delays $\{\hat{l}_k\}^{(c), j}$, the estimated responses can be expressed as $\hat{\bm{\gamma}}^{(c), j} = \tilde{\bm{\gamma}}^{(c), j}(\{\hat{l}_k\}^{(c), j})$. Note that if multiple targets are in the same direction, their delays need to be different. Using the relation 
\begin{equation}
  \hat{d}_k = \dfrac{v^l \hat{l}_k}{2},
\end{equation}
where $v^l$ denotes the speed of light, the range of the $k$-th target can be determined if the delay of echo signal $\hat{l}_k$ is decided.

To reduce the mis-detection probability caused by the noise and interference, a \emph{threshold detection} will be conducted for each target defined in hypothesis $U_j$ using the estimated target response $\hat{\bm{\gamma}}^{(c), j}$~\cite{skolnik2001Introduction}. To be specific, if $|\gamma^{(c), j}_k|$ is greater than a pre-determined positive threshold $\omega$, the $k$-th target is viewed to be present. Otherwise, the response $\gamma^{(c), j}_k$ is viewed to be caused by the noise and interference, and thus hypothesis $U_j$ will be rejected.

\nomenclature{$J_K$}{Number of hypotheses when the number of targets is $K$}%
\nomenclature{$\bm{y}^{(c)}$}{Received signals from the 1-st to the $c$-th cycle}%
\nomenclature{$p^{(c)}(\bm{y}^{(c)}|U_j)$}{Probability to receive $\bm{y}^{(c)}$ given hypothesis $U_j$}%
\nomenclature{$\{\hat{l}_k\}^{(c), j}$}{Estimated delays given hypothesis $U_j$ in the $c$-th cycle}%
\nomenclature{$\hat{\bm{\gamma}}^{(c), j}$}{Estimated responses given hypothesis $U_j$ in the $c$-th cycle}%
\nomenclature{$v^l$}{Speed of light}%
\nomenclature{$\omega$}{Threshold of the target response amplitude}%

\section{Joint Optimization Problem of Waveform and Phase Shifts}
\label{s_problem}

In this section, we formulate the waveform and RIS phase shift optimization problem in each cycle in Subsection~\ref{ss_pf}. Since the waveform and RIS phase shifts are coupled which are difficult to be simultaneously optimized, we decouple the formulated problem into three subproblems to solve it efficiently in Subsection~\ref{ss_pd}.

\subsection{Waveform and RIS Phase Shift Optimization Problem Formulation}
\label{ss_pf}

The aim to optimize the radar waveforms $\bm{w}$ and RIS phase shift vectors $\bm{s}^t$ and $\bm{s}^r$ in each cycle is to improve the detection performance. In this paper, we use the relative entropy as the evaluation criterion, which is shown to be an effective tool to study the performance of multiple hypotheses testing and is widely used for radar detection applications~\cite{goodman2007adaptive}. The relative entropy indicates the ``distance'' between the probability functions of two different hypotheses. When the ``distance'' is maximized, the hypotheses are more likely to be distinguished, and thus we can obtain a higher detection accuracy.

According to~\cite{wang2018multi}, we define the predicted distance between hypotheses $U_{j}$ and $U_{j'}$ in the $(c+1)$-th cycle as
\begin{align}
  d^{c+1}_{j, j'}(\mathcal{P}^{c+1}) =~& D(p^{c+1}(\bm{y}|U_{j}, \mathcal{P}^{c+1}), p^{c+1}(\bm{y}|U_{j'}, \mathcal{P}^{c+1})) \notag\\
  &+ D(p^{c+1}(\bm{y}|U_{j'}, \mathcal{P}^{c+1}), p^{c+1}(\bm{y}|U_{j}, \mathcal{P}^{c+1})),
\end{align}
where $\mathcal{P}^{c+1} = \{\bm{w}^{c+1}, \bm{s}^{t, c+1}, \bm{s}^{r, c+1}\}$ is the set of variables to be optimized in the $(c+1)$-th cycle. $p^{c+1}(\bm{y}|U_{j}, \mathcal{P}^{c+1})$ denotes the probability function given hypothesis $U_{j}$, variables $\mathcal{P}^{c+1}$, estimated target responses $\hat{\bm{\gamma}}^{c+1}$, and delays $\{\hat{l_k}\}^{c+1}$. Besides, $D(p^{c+1}(\bm{y}|U_{j}, \mathcal{P}^{c+1}), p^{c+1}(\bm{y}|U_{j'}, \mathcal{P}^{c+1}))$ is the relative entropy between probability functions $p^{c+1}(\bm{y}|U_{j}, \mathcal{P}^{c+1})$ and $p^{c+1}(\bm{y}|U_{j'}, \mathcal{P}^{c+1})$, where
\begin{align}
  &D(p^{c+1}(\bm{y}|U_{j}, \mathcal{P}^{c+1}), p^{c+1}(\bm{y}|U_{j'}, \mathcal{P}^{c+1})) \notag\\
  =& \int p^{c+1}(\bm{y}|U_{j}, \mathcal{P}^{c+1}) \log \dfrac{p^{c+1}(\bm{y}|U_{j}, \mathcal{P}^{c+1})}{p^{c+1}(\bm{y}|U_{j'}, \mathcal{P}^{c+1})} d\bm{y}.\label{def_D}
\end{align}

\nomenclature{$\mathcal{P}^{c+1}$}{Set of variables to be optimized in the $(c+1)$-th cycle}%
\nomenclature{$d^{c+1}_{j, j'}(\mathcal{P}^{c+1})$}{Predicted distance between hypotheses $U_j$ and $U_{j'}$ in the $(c+1)$-th cycle given $\mathcal{P}^{c+1}$}%

The objective of the optimization problem is to maximize the weighted sum of predicted distances between every two hypotheses. Therefore, the optimization problem in the $(c+1)$-th cycle can be formulated as\small
\begin{subequations}
  \begin{align}
    \text{P1:}\max_{\bm{w}^{c+1}, \bm{s}^{t, c+1}, \bm{s}^{r, c+1}}~& \sum^{J-1}_{j = 0}\sum^{J-1}_{j' = j+1}\beta^{c+1}_{j, j'} d^{c+1}_{j, j'}(\bm{w}^{c+1}, \bm{s}^{t, c+1}, \bm{s}^{r, c+1}),\label{p1_obj}\\
    s.t.~& ||\bm{w}^{c+1}||^2 = P_M, \label{p1_c1}\\
    & s^{t, c+1}_m \in \left\{ i\Delta s | i = 1, \cdots, N_s\right\},\forall m,\label{p1_c2}\\
    & s^{r, c+1}_m \in \left\{ i\Delta s | i = 1, \cdots, N_s\right\},\forall m,\label{p1_c3}
  \end{align}
\end{subequations}\normalsize
where $\beta^{c+1}_{j, j'}$ is the weighting factor for the predicted distance. Constraint (\ref{p1_c1}) is the power constraint for antennas~\cite{liu2020joint}, with constant $P_M$ being the maximum transmission power. Constraints (\ref{p1_c2}) and (\ref{p1_c3}) confine the available states of RIS elements~\cite{yu2020mimo}. The weighting factor is set as the prior probability product of two hypotheses, which is given by
\begin{equation}
  \beta^{c+1}_{j, j'} = p^{c+1}(U_j) p^{c+1}(U_{j'}).\label{def_beta}
\end{equation}
The intuition of (\ref{def_beta}) is that if the probabilities of two hypotheses are higher, the predicted distance of these two probabilities needs to have larger weight for better discrimination.

\nomenclature{$\beta^{c+1}_{j, j'}$}{Weighting factor for predicted distance $d^{c+1}_{j, j'}(\mathcal{P}^{c+1})$}%
\nomenclature{$P_M$}{Maximum transmission power}%

\subsection{Problem Decomposition}
\label{ss_pd}

It is difficult to tackle problem (P1) because variables $\bm{w}^{c+1}$, $\bm{s}^{t, c+1}$, and $\bm{s}^{r, c+1}$ are coupled in the objective function (\ref{p1_obj}). To solve problem (P1) efficiently, we decouple it into three subproblems, i.e., waveform optimization subproblem, RIS phase shift optimization subproblems for transmission and reception steps.

\subsubsection{Optimization Subproblem of Waveform}
\label{ssss}
Given the fixed RIS phase shift vectors $\bm{s}^{t}$, and $\bm{s}^{r}$, this subproblem is given by
\begin{subequations}
  \begin{align}
    \text{P}_w\text{:}\max_{\bm{w}^{c+1}}~& \sum^{J-1}_{j = 0}\sum^{J-1}_{j' = j+1}\beta^{c+1}_{j, j'} d^{c+1}_{j, j'}(\bm{w}^{c+1}, \bm{s}^{t}, \bm{s}^{r}),\label{p2_obj}\\
    s.t.~& \text{(\ref{p1_c1})}.\notag
  \end{align}
\end{subequations}

\subsubsection{Optimization Subproblem of RIS Phase Shift Vector in the Transmission Step}

In this subproblem, the RIS phase shift vector $\bm{s}^{t, c+1}$ is optimized given $\bm{w}$ and $\bm{s}^{r}$, which can be written as
\begin{subequations}
  \begin{align}
    \text{P}_t\text{:}\max_{\bm{s}^{t, c+1}}~& \sum^{J-1}_{j = 0}\sum^{J-1}_{j' = j+1}\beta^{c+1}_{j, j'} d^{c+1}_{j, j'}(\bm{w}, \bm{s}^{t, c+1}, \bm{s}^{r}),\label{p3_obj}\\
    s.t.~& \text{(\ref{p1_c2})}.\notag
  \end{align}
\end{subequations}

\subsubsection{Optimization Subproblem of RIS Phase Shift Vector in the Reception Step}

Similarly, the optimization subproblem of the RIS phase shift vector in the reception step can be formulated as
\begin{subequations}
  \begin{align}
    \text{P}_r\text{:}\max_{\bm{s}^{r, c+1}}~& \sum^{J-1}_{j = 0}\sum^{J-1}_{j' = j+1}\beta^{c+1}_{j, j'} d^{c+1}_{j, j'}(\bm{w}, \bm{s}^{t}, \bm{s}^{r, c+1}),\label{p3_obj}\\
    s.t.~& \text{(\ref{p1_c3})}.\notag
  \end{align}
\end{subequations}

\section{Algorithm Design}
\label{s_algorithm}

In this section, we propose a waveform and phase shift optimization (WPSO) algorithm, which solves problem (P1) by iteratively solving the aforementioned three subproblems. The techniques to solve these three subproblems are introduced first in Subsections~\ref{ss_ow} and \ref{ss_ops}, and then the description of the overall WPSO algorithm is presented in Subsection~\ref{ss_oa}. The superscript $c+1$ for cycles are omitted for brevity in the following part of this paper.

\subsection{Optimization of Waveform}
\label{ss_ow}

It is challenging to solve subproblem ($\text{P}_w$) due to the complicated complicated expression of the predicted distance $d(U_j, U_{j'}|\mathcal{P})$ in the objective function~(\ref{p1_obj}). In Proposition 1, a simplified expression of $d(U_j, U_{j'}|\mathcal{P})$ is provided as follows:

\newtheorem{proposition}{\bf Proposition}

\begin{proposition}
  \label{pro_d}
  The predicted distance $d(U_j, U_{j'}|\mathcal{P})$ between hypotheses $U_j$ and $U_{j'}$ given $\mathcal{P}$ can be expressed as
  \begin{equation}
    d(U_j, U_{j'}|\mathcal{P}) = \dfrac{1}{\sigma^2}\text{Re}\left(\text{tr}(\bm{w}^{\text{H}} \bm{Z}'(U_j, U_{j'}, \bm{s}^t, \bm{s}^r)\bm{w})\right),
  \end{equation}
  where \small
  \begin{align}
  \bm{Z}'(U_j, U_{j'}, \bm{s}^t, \bm{s}^r) = &\bm{Z}''(U_j, U_{j}, \bm{s}^t, \bm{s}^r) + \bm{Z}''(U_{j'}, U_{j'}, \bm{s}^t, \bm{s}^r) \notag\\
  &- 2\bm{Z}''(U_j, U_{j'}, \bm{s}^t, \bm{s}^r),\\
  \bm{Z}''(U_j, U_{j'}, \bm{s}^t, \bm{s}^r) =& \sum^{N(U_j)}_{i=1}\sum^{N(U_{j'})}_{i'=1} \hat{\gamma}^{\text{H}}_i(U_j)\hat{\gamma}_{i'}(U_{j'})\bm{Q}^{\text{H}}_i(U_j, \bm{s}^t, \bm{s}^r) \notag\\
  &\times \bm{Q}_{i'}(U_{j'}, \bm{s}^t, \bm{s}^r).
  \end{align}\normalsize
\end{proposition}
\begin{IEEEproof}
  See Appendix \ref{proof_d}.
\end{IEEEproof}

Therefore, problem ($\text{P}_w$) can be reformulated as 
\begin{subequations}
  \begin{align}
    \text{P}'_w\text{:}\max_{\bm{w}}~& \text{Re}\left(\text{tr}(\bm{w}^{\text{H}} \bm{Z}(\bm{s}^t, \bm{s}^r) \bm{w})\right),\label{p_w2_obj}\\
    s.t.~& \text{tr}(\bm{w}^{\text{H}}\bm{w}) = P_M, \label{p_w2_c1}
  \end{align}
\end{subequations}
where $\bm{Z}(\bm{s}^t, \bm{s}^r) = \sum^{J-1}_{j = 0}\sum^{J-1}_{j' = j+1} \dfrac{\beta_{j, j'}}{\sigma^2}\bm{Z}'(U_j, U_{j'}, \bm{s}^t, \bm{s}^r)$, and constraint (\ref{p_w2_c1}) corresponds to constraint (\ref{p1_c1}). Since problem ($\text{P}'_w$) is a quadratically constrained quadratic program (QCQP), we can use the semidefinite relaxation (SDR) technique~\cite{luo2010semidefinite} to efficiently solve it. Specifically, problem ($\text{P}'_w$) is equivalent to the following problem:
\begin{subequations}
  \begin{align}
    \text{P}_{\bm{X}_w}\text{:}\max_{\bm{X}_w}~& \text{Re}\left(\text{tr}(\bm{X}_w \bm{Z}(\bm{s}^t, \bm{s}^r) )\right),\label{p_x_w_obj}\\
    s.t.~& \text{tr}(\bm{X}_w) = P_M, \label{p_x_w_c1}\\
    & \bm{X}_w \succcurlyeq 0,\\
    & \bm{X}_w \in \mathbb{H}^{NL\times NL},\\
    & \text{rank}(\bm{X}_w) = 1, \label{p_x_w_c2}
  \end{align}
\end{subequations}
where $\bm{X}_w = \bm{w}\bm{w}^{\text{H}}$, and $\bm{X}_w \succcurlyeq 0$ indicates that $\bm{X}_w$ is a positive semidefinite matrix. To solve problem ($\text{P}_{\bm{X}_w}$), the rank constraint in~(\ref{p_x_w_c2}) is first dropped to derive a relaxed version of problem ($\text{P}_{\bm{X}_w}$), denoted by ($\text{P}'_{\bm{X}_w}$), which can be efficiently solved using existing optimization techniques~\cite{ye1997interior}. Then, the randomization method in~\cite{luo2010semidefinite} can be adopted to convert the solution of problem ($\text{P}'_{\bm{X}_w}$) to a feasible solution of problem ($\text{P}_{\bm{X}_w}$), which is also the solution of problem ($\text{P}_w$).

\subsection{Optimization of RIS Phase Shift Vector in the Transmission and Reception Steps}
\label{ss_ops}

Due to the similar structures of problems ($\text{P}_t$) and ($\text{P}_r$), these two problems can be solved by the same optimization methods. For brevity, in this subsection we only show how to solve problem ($\text{P}_t$).

To tackle problem ($\text{P}_t$) where phase shift $\bm{s}^t$ is optimized, in Proposition~\ref{pro_d2}, we provide another simplified expression of distance $d(U_j, U_{j'}|\mathcal{P})$ which separates variable $\bm{r}(\bm{s}^t)$ from other parameters as follows:

\begin{proposition}
  \label{pro_d2}
  The predicted distance $d(U_j, U_{j'}|\mathcal{P})$ between hypotheses $U_j$ and $U_{j'}$ given $\mathcal{P}$ can be expressed as\small
  \begin{align}
    d(U_j, U_{j'}|\mathcal{P}) =& \dfrac{1}{\sigma^2}\text{Re}\Big(\bm{r}^{\text{H}}(\bm{s}^t) \bm{Z}'(U_j, U_{j'}, \bm{w}, \bm{s}^r)\bm{r}(\bm{s}^t) \notag\\
    &+ \bm{r}^{\text{H}}(\bm{s}^t) \bm{z}'_1(U_j, U_{j'}, \bm{w}, \bm{s}^r) \notag\\
    &+ \bm{z}'_2(U_j, U_{j'}, \bm{w}, \bm{s}^r) \bm{r}(\bm{s}^t) + z'_3 (U_j, U_{j'}) \Big),
  \end{align}\normalsize
  where parameter $\bm{Z}'(U_j, U_{j'}, \bm{w}, \bm{s}^r)$, $\bm{z}'_1(U_j, U_{j'}, \bm{w}, \bm{s}^r)$, $\bm{z}'_2(U_j, U_{j'}, \bm{w}, \bm{s}^r)$, and $z'_3 (U_j, U_{j'})$ are defined in Appendix~\ref{proof_d2}.
\end{proposition}
\begin{IEEEproof}
  See Appendix~\ref{proof_d2}.
\end{IEEEproof}

Consequently, we can reformulated problem ($\text{P}_t$) as 
\begin{subequations}
  \begin{align}
    \text{P}'_t\text{:}\max_{\bm{r}}~& \text{Re}\!\left(\bm{r}^{\text{H}} \bm{Z}(\bm{w}, \bm{s}^r) \bm{r} \!+\! \bm{r}^{\text{H}}\bm{z}_1(\bm{w}, \bm{s}^r) \!+\! \bm{z}_2(\bm{w}, \bm{s}^r)\bm{r} \!+\! z_3\right),\label{p_t2_obj}\\
    s.t.~& r_m \in \left\{\eta e^{ji\Delta s}| i = 1, \cdots, N_s\right\}, \forall m, \label{p_t2_c1}
  \end{align}
\end{subequations}
where 
\begin{align}
  \bm{Z}(\bm{w}, \bm{s}^r) =& \sum^{J-1}_{j = 0}\sum^{J-1}_{j' = j+1} \dfrac{\beta_{j, j'}}{\sigma^2}\bm{Z}'(U_j, U_{j'}, \bm{w}, \bm{s}^r),\\
  \bm{z}_1(\bm{w}, \bm{s}^r) =& \sum^{J-1}_{j = 0}\sum^{J-1}_{j' = j+1} \dfrac{\beta_{j, j'}}{\sigma^2}\bm{z}'_1(U_j, U_{j'}, \bm{w}, \bm{s}^r),\\
  \bm{z}_2(\bm{w}, \bm{s}^r) =& \sum^{J-1}_{j = 0}\sum^{J-1}_{j' = j+1} \dfrac{\beta_{j, j'}}{\sigma^2}\bm{z}'_2(U_j, U_{j'}, \bm{w}, \bm{s}^r),\\
  z_3 =& \sum^{J-1}_{j = 0}\sum^{J-1}_{j' = j+1} \dfrac{\beta_{j, j'}}{\sigma^2}z'_3(U_j, U_{j'}),
\end{align}
And constraint (\ref{p_t2_c1}) corresponds to constraint (\ref{p1_c2}). To efficiently solve problem ($\text{P}'_t$), we first relax the discrete phase shifts to continuous ones, and obtain the following problem:\small
\begin{subequations}
  \begin{align}
    \text{P}''_t\text{:}\max_{\bm{r}}~& \text{Re}\left(\bm{r}^{\text{H}} \bm{Z}(\bm{w}, \bm{s}^r) \bm{r} + \bm{r}^{\text{H}}\bm{z}_1(\bm{w}, \bm{s}^r) + \bm{z}_2(\bm{w}, \bm{s}^r)\bm{r} + z_3\right),\label{p_t3_obj}\\
    s.t.~& |r_m|^2 = \eta^2, m = 1, \cdots, M.\label{p_t3_c1}
  \end{align}
\end{subequations}\normalsize
Since problem ($\text{P}''_t$) is also a QCQP, the SDR technique can also be applied to solve problem ($\text{P}''_t$). Let $\bm{r}''$ denote the solution of problem ($\text{P}''_t$). Next, all the elements in vector $\bm{r}''$ are quantized to the nearest values in the set $\left\{\eta e^{ji\Delta s}| i = 1, \cdots, N_s\right\}$ to obtain $\bm{r}'$. Vector $\bm{r}'$ is a feasible solution of problem ($\text{P}'_t$) because it satisfies constraint~(\ref{p_t2_c1}). Finally, the solution of problem~($\text{P}_t$) can be derived from $\bm{r}'$ using the relation in~(\ref{def_r_m}).

\subsection{Joint Optimization Algorithm}
\label{ss_oa}

\begin{figure}[!t]
  \centering
  \includegraphics[width=3in]{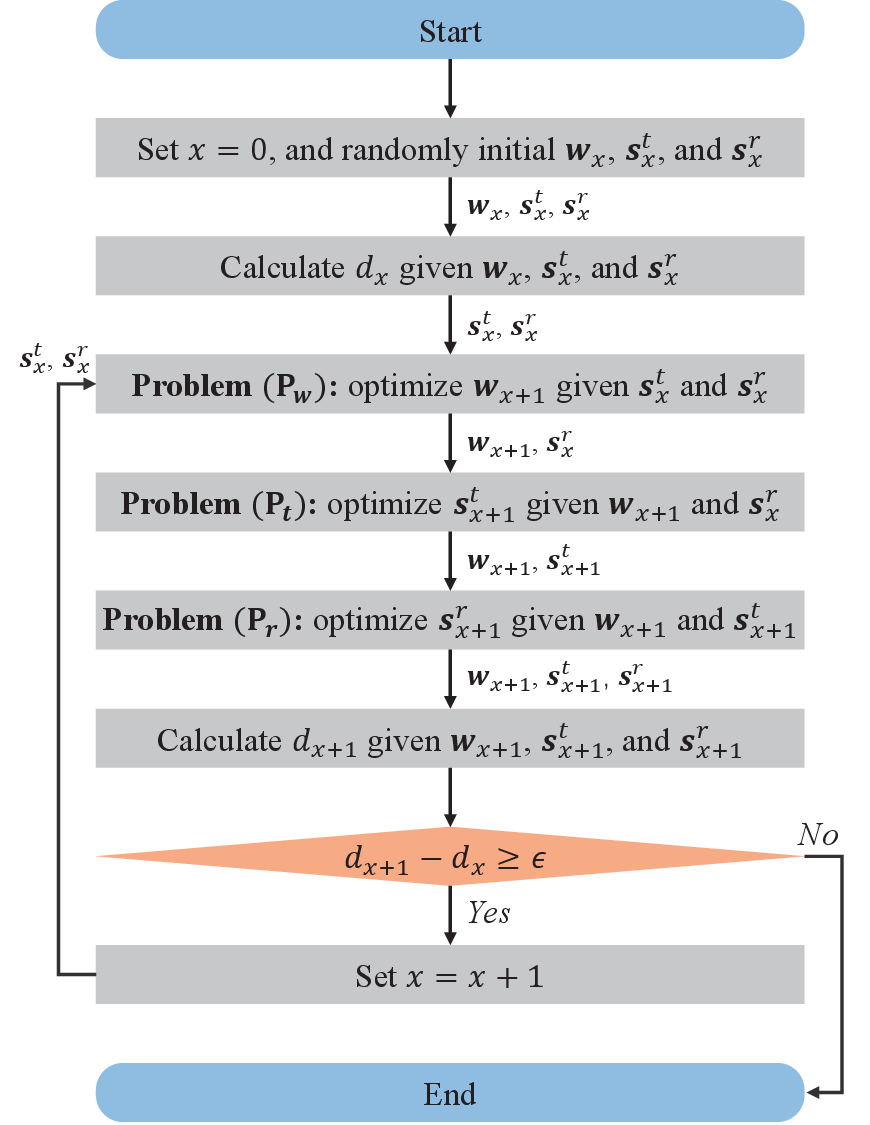}
  \textcolor{black}{\caption{Flow chart of the waveform and phase shift optimization algorithm.}}
\end{figure}

\textcolor{black}{As shown in Fig.~5, the index of iteration $x$ is set as $0$ at the beginning of the WPSO algorithm. Besides, we randomly initial variables $\bm{w}_{x}$, $\bm{s}^{t}_{x}$, and $\bm{s}^{r}_{x}$ within the feasible region of problem ($\text{P1}$). Then, the value of the objective function~(\ref{p1_obj}) $d_x$ given $\bm{w}_{x}$, $\bm{s}^{t}_{x}$, and $\bm{s}^{r}_{x}$ is calculated. Following that, we optimize the radar waveform and RIS phase shifts in an iterative manner by solving the three subproblems. Specifically, in the $x$-th iteration, the WPSO algorithm first solves subproblem ($\text{P}_w$) to obtain $\bm{w}_{x+1}$ given $\bm{s}^{t}_{x}$ and $\bm{s}^{r}_{x}$. Then, subproblem ($\text{P}_t$) is solved to derive $\bm{s}^{t}_{x+1}$ given $\bm{w}_{x+1}$ and $\bm{s}^{r}_{x}$. The variable $\bm{s}^{r}_{x+1}$ is finally derived by solving subproblem ($\text{P}_r$) given $\bm{w}_{x+1}$ and $\bm{s}^{t}_{x+1}$. The algorithm will terminate if the value difference of the objective function (\ref{p1_obj}) in two adjacent iterations is smaller than a predefined threshold~$\epsilon$.}

\section{Performance Analysis}
\label{s_analysis}

In this section, we we first analyse the convergence and the complexity of the proposed WPSO algorithm in Subsection~\ref{ss_cc}, and then discuss the detection performance of the proposed scheme in Subsection~\ref{ss_pc}.

\subsection{Convergence and Complexity}
\label{ss_cc}

\subsubsection{Convergence}
\label{ss_convergence}

In each iteration of the WPSO algorithm, the subproblems ($\text{P}_w$), ($\text{P}_t$), and ($\text{P}_r$) are solved sequentially using the interior-point algorithm, which is guaranteed to converge~\cite{ye1997interior}. Hence, the WPSO algorithm will converge if the number of iterations of the WPSO algorithm is limited. Since the objective function is increased by at least $\epsilon$ in each iteration and the objective function has an upper bound $d^u$ which is provided in Proposition~\ref{pro_obj_upper}, the number of iterations is bounded and the convergence of the WPSO algorithm is guaranteed.

\begin{proposition}
  \label{pro_obj_upper}
  An upper bound of the objective function~(\ref{p1_obj}) is given by
  \begin{equation}
    d^u = \dfrac{J(J-1)P_M}{2}.
  \end{equation}
\end{proposition}
\begin{IEEEproof}
  See Appendix \ref{proof_obj_upper}.
\end{IEEEproof}

\nomenclature{$d^u$}{An upper bound of~(\ref{p1_obj})}%

\subsubsection{Complexity}
\label{sss_complexity_analysis}

In each iteration of the WPSO algorithm, the SDR technique is utilized for 3 times to solve the three subproblems. According to~\cite{luo2010semidefinite}, the complexity of the SDR techniques for the subproblems $(\text{P}_w)$, $(\text{P}_t)$, and $(\text{P}_r)$ are $O((NL)^{4.5}\log(1/\alpha))$, $O((M+1)^{4.5}\log(1/\alpha))$, and $O((M+1)^{4.5}\log(1/\alpha))$, respectively, where $\alpha$ is the solution accuracy of the interior-point algorithm. Thus, the complexity of each iteration of the WPSO algorithm is $O(((NL)^{4.5}+(M+1)^{4.5})\log(1/\alpha))$.

\nomenclature{$\alpha$}{Solution accuracy of the interior-point algorithm}%

Based on~(\ref{app_A_1}), the objective function~(\ref{p1_obj}) is nonnegative. According to the discussion in Section~\ref{ss_convergence}, the maximum number of iterations of the WPSO algorithm is $\lceil d^u/\epsilon\rceil$. Therefore, the complexity of the WPSO algorithm is $O(((NL)^{4.5}+(M+1)^{4.5})\log(1/\alpha)\lceil d^u/\epsilon\rceil)$.

\subsection{Detection Performance Analysis}
\label{ss_pc}

In this subsection, we compare the detection performance of MetaRadar with traditional MIMO radar. For simplicity, we consider the case where $K = 1$ and $I = 1$. There are $2$ hypotheses in this case denoted by $U_0$ and $U_1$. Based on~(\ref{app_A_1}), the predicted distance between hypotheses $U_0$ and $U_1$ of MetaRadar in this case is given by
\begin{align}
  d(U_1, U_{0}|\mathcal{P}) =& \dfrac{1}{\sigma^2} ||\overline{\bm{y}}(U_{1}, \mathcal{P}) - \overline{\bm{y}}(U_{0}, \mathcal{P})||^2\notag\\
  =& \dfrac{1}{\sigma^2} ||\gamma (\bm{b}^{\text{T}}_1(\bm{s}^{r}) + \bm{\xi}^{\text{T}}_1) (\bm{b}_1(\bm{s}^{t}) + \bm{\xi}_1) \bm{W} ||^2\notag\\
  =& \dfrac{|\gamma|^2}{\sigma^2} ||\bm{b}_1(\bm{s}^{r}) + \bm{\xi}_1||^2 ||\bm{b}_1(\bm{s}^{t}) + \bm{\xi}_1||^2||\bm{W}||^2\notag\\
  =& \dfrac{|\gamma|^2 P_M}{\sigma^2} ||\bm{b}_1(\bm{s}^{r}) + \bm{\xi}_1||^2 ||\bm{b}_1(\bm{s}^{t}) + \bm{\xi}_1||^2.\label{d_smart_radar}
\end{align}

As for the MIMO radar, its received signals can be expressed as~\cite{wang2018multi} 
\begin{equation}
  \bm{Y}_{MIMO} = \sum^K_{k=1} \gamma_k \bm{\xi}^{\text{T}}_k \bm{\xi}_k \bm{W} \bm{J}_k + \bm{V},
\end{equation}
Similar to the proof in Appendix~\ref{proof_d}, the predicted distance of traditional MIMO radar between hypotheses $U_0$ and $U_1$ when $K = 1$ is given by
\begin{align}
  d_{MIMO}(U_1, U_{0}|\bm{W}) =& \dfrac{1}{\sigma^2}||\gamma \bm{\xi}^{\text{T}}_1 \bm{\xi}_1 \bm{W} \bm{J}_1||^2 \notag\\
  =& \dfrac{|\gamma|^2 P_M}{\sigma^2}||\bm{\xi}_1||^4\notag\\
  =& \dfrac{(N|\gamma|G^A G^A_P(\theta_1, \varphi_1))^2 P_M}{\sigma^2}.\label{d_mimo_radar}
\end{align}

It can be observed from~(\ref{d_smart_radar}) and (\ref{d_mimo_radar}) that the predicted distances of both MetaRadar and MIMO radar are positively related to $|\gamma|^2$ and the power of waveform $P_M$, and are negatively related to the variance of the residual term $\sigma^2$. Besides, we can conclude that if $||\bm{b}_1(\bm{s}^{r}) + \bm{\xi}_1||^2 ||\bm{b}_1(\bm{s}^{t}) + \bm{\xi}_1||^2 > (NG^A G^A_P(\theta_1, \varphi_1))^2$, the predicted distance of the MetaRadar is larger than that of the traditional MIMO radar. In this circumstance, hypotheses $U_0$ and $U_1$ can be better distinguished by the MetaRadar, indicating that MetaRadar has a higher detection accuracy.

Since $\bm{b}_1(\bm{s}^{t}) + \bm{\xi}_1$ and $\bm{b}_1(\bm{s}^{r}) + \bm{\xi}_1$ denote the channel gain from the antenna array to the direction of the target and vice versa, respectively, $||\bm{b}_1(\bm{s}^{r}) + \bm{\xi}_1||^2 ||\bm{b}_1(\bm{s}^{t}) + \bm{\xi}_1||^2$ can be viewed as the power gain of the MetaRadar. As the power gain of the MetaRadar is determined by the phase shift vectors and the configuration of the MetaRadar, in the following, we first analyse the optimal phase shift vectors which maximize the power gain of the MetaRadar given the RIS configuration, and then discuss the the relationship between the configuration of the MetaRadar and the power gain when RIS phase shift vectors are optimized.

\subsubsection{Optimal RIS Phase Shift Vectors}
The maximum power gain given the configuration of the MetaRadar when $N = 1$ is provided in Proposition~\ref{pro_maximum} as follows:
\begin{proposition}
  \label{pro_maximum}
  Assume that all the RIS elements have continuous phase shifts, the maximum power gain given the configuration of the MetaRadar when $N = 1$ is\small
  \begin{align}
    B \!=\!& ||\bm{b}_1(\bm{s}^{r, *}) \!+\! \bm{\xi}_1||^2 ||\bm{b}_1(\bm{s}^{t, *}) \!+\! \bm{\xi}_1||^2 \notag\\
    =& \left(G^A \right)^2\!\!\Bigg(\sum^M_{m=1} \!\dfrac{\rho\sqrt{G^A_P(\theta^r_{1, m}, \varphi^r_{1, m}) G^R_P(\theta^r_{1, m}, \varphi^r_{1, m}) }}{l_{1, m}} \notag\\
    &\!+\! \sqrt{G^A_P(\theta_1, \varphi_1)}\Bigg)^4,\label{pro_maximum_B}
  \end{align}\normalsize
  where 
  \begin{equation}
    s^{t, *}_m = s^{r, *}_m = \text{mod}(-\bm{e}_1\bm{p}^e_m - \bm{e}_1 \bm{p}^a_1 - \dfrac{2\pi l_{1, m}}{\lambda}, 2\pi), \forall m,\label{pro_maximum_s}
  \end{equation}
  and
  \begin{align}
    \rho = \dfrac{\eta \sqrt{G^R S^e G^R_P(\theta_1, \varphi_1)}}{\sqrt{4\pi}}.
  \end{align}

\end{proposition}
\begin{IEEEproof}
The term $||\bm{b}_1(\bm{s}) + \bm{\xi}_1||^2$ can be expressed as\small
\begin{align}
  &||\bm{b}_1(\bm{s}) + \bm{\xi}_1||^2 \notag\\
  =& ||\bm{a}_1 \bm{R}(\bm{s})\bm{H} + \bm{\xi}_1||^2\notag\\  
  =& \left|\sum^M_{m=1} a_{1, m}r_m(s_m) h_{m, 1} + \xi_{1, 1}\right|^2 \notag\\
  =&  G^A\! \Bigg|\sum^M_{m=1}\! \dfrac{\rho\sqrt{G^A_P(\theta^r_{1, m}, \varphi^r_{1, m}) G^R_P(\theta^r_{1, m}, \varphi^r_{1, m}) }}{l_{1, m}} e^{-j(\bm{e}_1 \bm{p}^e_m + s_m+ 2\pi l_{1, m}/\lambda)} \notag\\
  &+ \sqrt{G^A_P(\theta_1, \varphi_1)}e^{j\bm{e}_1 \bm{p}^a_1}\Bigg|^2\notag\\
  \le& G^A\! \left(\sum^M_{m=1}\! \dfrac{\rho\sqrt{G^A_P(\theta^r_{1, m}, \varphi^r_{1, m}) G^R_P(\theta^r_{1, m}, \varphi^r_{1, m})}}{l_{1, m}} \!+\! \sqrt{G^A_P(\theta_1, \varphi_1)}\right)^2,
\end{align}\normalsize
where 
\begin{align}
  \rho = \dfrac{\eta \sqrt{G^R S^e G^R_P(\theta_1, \varphi_1)}}{\sqrt{4\pi}},
\end{align}
and $||\bm{b}_1(\bm{s}) + \bm{\xi}_1||^2$ is maximized when
\begin{align}
  s_m = \text{mod}(-\bm{e}_1\bm{p}^e_m - \bm{e}_1 \bm{p}^a_1 - \dfrac{2\pi l_{1, m}}{\lambda}, 2\pi).
\end{align}

Consequently, the maximum power gain of the MetaRadar when $N = 1$ is (\ref{pro_maximum_B}), and the optimal RIS phase shift vectors satisfy (\ref{pro_maximum_s}).
\end{IEEEproof}

\subsubsection{Configuration of the MetaRadar}
\label{sss_configuration}

\begin{figure}[!t]
  \centering
  \includegraphics[width=2.5in]{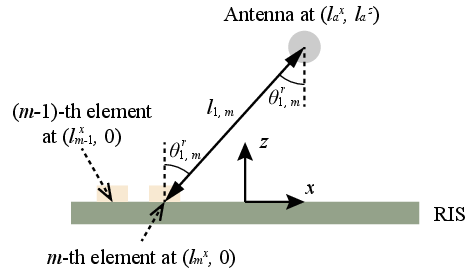}
  \caption{Top view of the MetaRadar.}
  \label{f_top_view}
\end{figure}

The configuration of the MetaRadar when $N = 1$ is shown in Fig.~\ref{f_top_view}. For simplicity, we consider a two-dimensional~(2D) configuration, and the extension to 3D configuration is feasible. The RIS is parallel to the $x$-axis and its center is at the origin $(0, 0)$. In the following, we discuss how the number of RIS elements and the antenna position affect the power gain $B$.

\textbf{Number of RIS elements:} The power gain is positively related to the number of RIS elements. This is because if we add an element at the edge of the RIS and keep the positions of existing $M$ elements fixed, according to~(\ref{pro_maximum_B}), the power gain will increase because the term $\sqrt{G^A_P(\theta^r_{1, m}) G^R_P(\theta^r_{1, m}) }/l_{1, M+1} > 0$. This indicates that an RIS with a larger size can provide a higher detection accuracy.

\textbf{Antenna position:} In the following proposition, we show that the antenna should be placed near the $y$-axis to promote the detection performance.

\begin{proposition}
  \label{pro_opt_l_x_a}
  Suppose the antenna is isotropic ($G^A = 1$ and $G^A_P(\theta, \varphi) = 1, \forall \theta, \varphi$). The power gain of the MetaRadar where the antenna is close to the $z$-axis is higher than that of the MetaRadar where the antenna is far away from the $z$-axis given the number of the RIS elements $M$ and the $z$-coordinate of the antenna $l^z_a$. Specifically, if $l^x_a \notin [-l^e/2, l^e/2]$, there exists a $l' \in [-l^e/2, l^e/2]$ with higher power gain than that of $l^x_a$, where $l^e$ is the distance between two adjacent RIS elements. 
\end{proposition}
\begin{IEEEproof}
  See Appendix \ref{proof_opt_l_x_a}.
\end{IEEEproof}

When $l^z_a \gg l^e/2$, the power gain when $l^x_a$ is optimized is given by Proposition~\ref{pro_max_b_l_x_a}:

\begin{proposition}
  \label{pro_max_b_l_x_a}
  When $l^z_a \gg l^e/2$, the power gain when $l^x_a$ is optimized can be expressed as
  \begin{align}
    B(l^*, l^z_a) =& \left(\sum^M_{m=1} \dfrac{\rho(l^z_a)^{1.5}}{((l^z_a)^2 + ((M+1)l^e/2 - m l^e)^2)^{1.25}} + 1\right)^4.\label{pro_max_b_l_x_a_1}
  \end{align}

\end{proposition}
\begin{IEEEproof}
  If $l^z_a \gg l_e/2$, we have $l^z_a \gg |l'|$, which indicates that the term $l'$ in the denominator in~(\ref{app_d_3}) can be omitted. Therefore, the maximum power gain can be expressed as~(\ref{pro_max_b_l_x_a_1}) based on~(\ref{app_d_3}).
\end{IEEEproof}

As for the $z$-coordinate of the antenna $l^z_a$, since both the numerator $(l^z_a)^{1.5}$ and the denominator $((l^z_a)^2 + ((M+1)l^e/2 - m l^e)^2)^{1.25}$ in~(\ref{pro_max_b_l_x_a_1}) increase with $l^z_a$, its relationship with the power gain is much more complicated which makes it difficult to provide a closed-form expression of the optimal $z$-coordinate of the antenna.

\begin{table}[!t]
  \caption{Simulation Parameters}
  \label{t_simulation}
  \centering
  \begin{tabular}{|l|l|}
    \hline
    \textbf{Parameters} & \textbf{Values}\\
    \hline
    \hline Position of the RIS's center $\bm{p}^R$ & $(0, 0, 0)$\\
    \hline Number of RIS elements $M$ & $64$\\
    \hline Area of an RIS element $S^e$ & $(\lambda/2)^2$\\
    \hline Position of the antenna array's center $\bm{p}^A$ & $(0, 3\lambda, 0)$\\
    \hline Number of antennas $N$ & $4$\\
    \hline Separation between adjacent antennas $d^s$ & $\lambda/2$\\
    \hline Gain of an antenna $G^A$ & $1$\\
    \hline Normalized radiation pattern of an antenna $G^A_P(\theta, \varphi)$ & $1$\\
    \hline Maximum transmission power of the antenna array $P_M$ & $12$W\\
    \hline Variance of the residual term $\sigma^2$ & $-50$dBw\\
    \hline Number of snapshots of the radar waveform $L$ & $10$\\
    \hline Number of snapshots of the received signal $L_R$ & $15$\\
    \hline Number of angular grids $I$ & $4$\\
    \hline Number of targets $K$ & $2$\\
    \hline Amplitude of target response $|\gamma_k|$ & $-40$dB\\
    \hline
  \end{tabular}
  \vspace{-6mm}
\end{table}

\nomenclature{$\bm{p}^R$}{Position of the RIS's center}%
\nomenclature{$S^e$}{Area of an RIS element}%
\nomenclature{$\bm{p}^A$}{Position of the antenna array's center}%
\nomenclature{$d^s$}{Separation between adjacent antennas}%
\nomenclature{$(\cdot)^*$}{Conjugate operator}%
\nomenclature{$(\cdot)^{\text{T}}$}{Transpose operator}%
\nomenclature{$(\cdot)^{\text{H}}$}{Conjugate transpose operator}%
\nomenclature{$\otimes$}{Kronecker product}%
\nomenclature{$\mathbb{C}^{M \times N}$}{Set of all complex $M \times N$ matrices}%
\nomenclature{$\mathbb{H}^{M \times N}$}{Set of all Hermitian $M \times N$ matrices}%
\nomenclature{$\bm{0}^{M \times N}$}{Matrix with all the elements being $0$ and size $M \times N$}%
\nomenclature{$\bm{I}^{M \times N}$}{Identity matrix with size $M \times N$}%
\nomenclature{$\mathbb{E}(\cdot)$}{Stochastic expectation}%
\nomenclature{$|| \bm{\beta} ||$}{Euclidean norm of vector $\bm{\beta}$}%
\nomenclature{tr$(\cdot)$}{Trace operator}%
\nomenclature{vec$(\cdot)$}{Vector operator}%
\nomenclature{Re$(\cdot)$}{Real part of a complex variable}%

\section{Simulation Results}
\label{s_simulation}

In this section, the performance of the MetaRadar is provided. The configuration of the MetaRadar is shown in Fig.~\ref{f_model}, and the simulation parameters are listed in Table~\ref{t_simulation}. The RIS is located at the plane $z = 0$, and the position of its center is $(0, 0, 0)$. The RIS contains $64$ elements. Each element has $8$ different phase shifts, and the area of an element is $(\lambda/2)^2$~\cite{zhang2021metalocalization}. The antennas are arranged as a $2\times 2$ array, and the separation between adjacent antennas is $\lambda/2$~\cite{deng2021reconfigurable}. The center of the antenna array is at $(0, 3\lambda, 0)$. Each antenna is assumed to be isotropic. That is, $G^A = 1$ and $G^A(\theta, \varphi) = 1, \forall \theta, \varphi$. The maximum transmission power of the antenna array is set as $12$W. The variance of the residual term $\sigma^2$ is $-50$dBw, and threshold $\omega = \sigma / 60$. The number of snapshots of the radar waveform is $10$, and the number of snapshots of the received signal is $15$. The angular range of interest is $\theta = \pi/6$, $\varphi \in [0, 2\pi)$, and the range $[0, 2\pi)$ is uniformly divided into $4$ grids. We assume that the number of targets is $2$. The two targets are located at $(10\Delta d, \pi/6, \pi/4)$ and $(15\Delta d, \pi/6, 3\pi/4)$. Here, $\Delta d$ denotes the length of a range cell, which is normalized as $1$. The correct hypothesis is denoted by $U_{j^*}$. The amplitude of target response $|\gamma_k|$ is $-40$dB, and the phase of the target response is randomly selected in $[0, 2\pi)$ in each Monte Carlo run.

\vspace{-4mm}
\subsection{Performance Comparison}
\vspace{-2mm}

In this subsection, we compare the performance of the proposed scheme with those of the random scheme and traditional MIMO radar scheme based on the relative entropy criterion~\cite{wang2018multi}. In the random scheme, the radar waveforms have fixed envelope and random phases. The phase shift vectors of the RIS in the transmission and reception steps are also randomly generated.

\begin{figure*}[!t]
  \centering
  \subfloat[]{
    \includegraphics[height=1.5in]{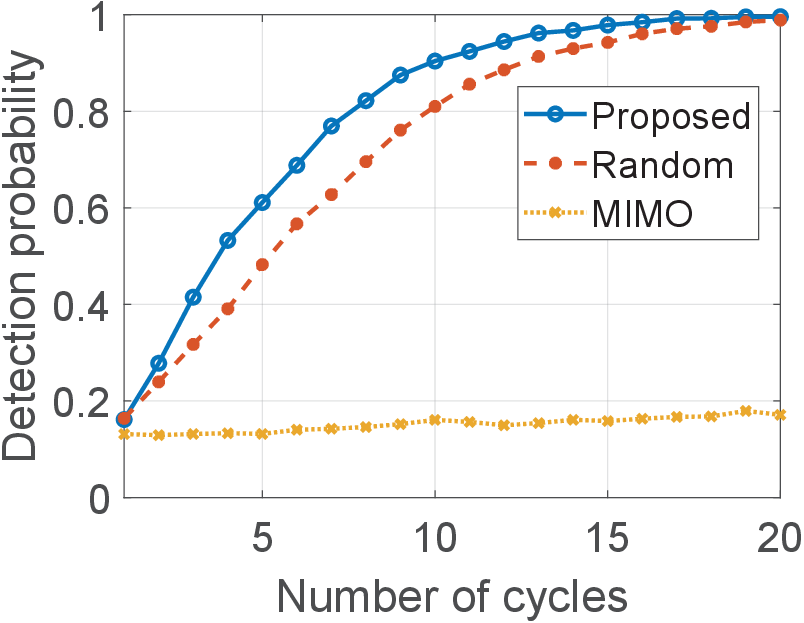}
  }
  \hspace{0.05in}
  \subfloat[]{
    \includegraphics[height=1.5in]{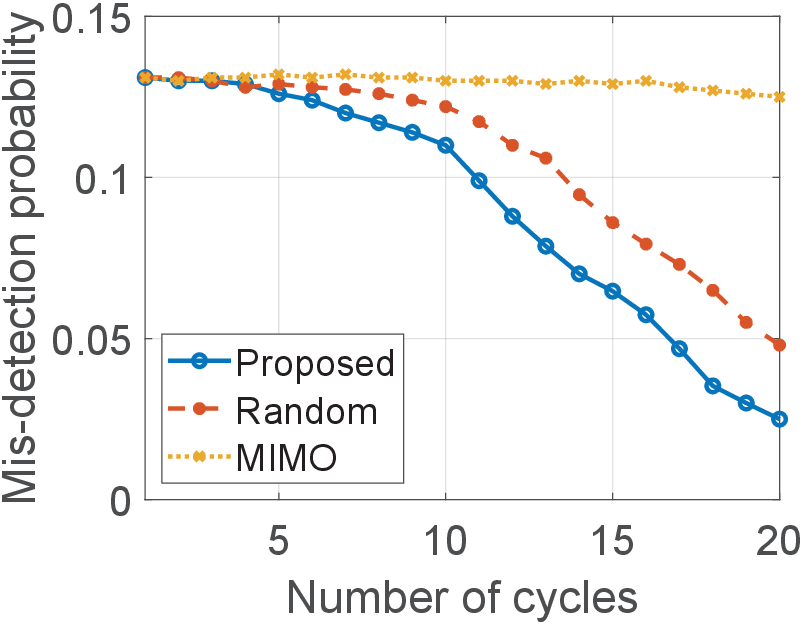}
  }
  \hspace{0.05in}
  \subfloat[]{
      \includegraphics[height=1.5in]{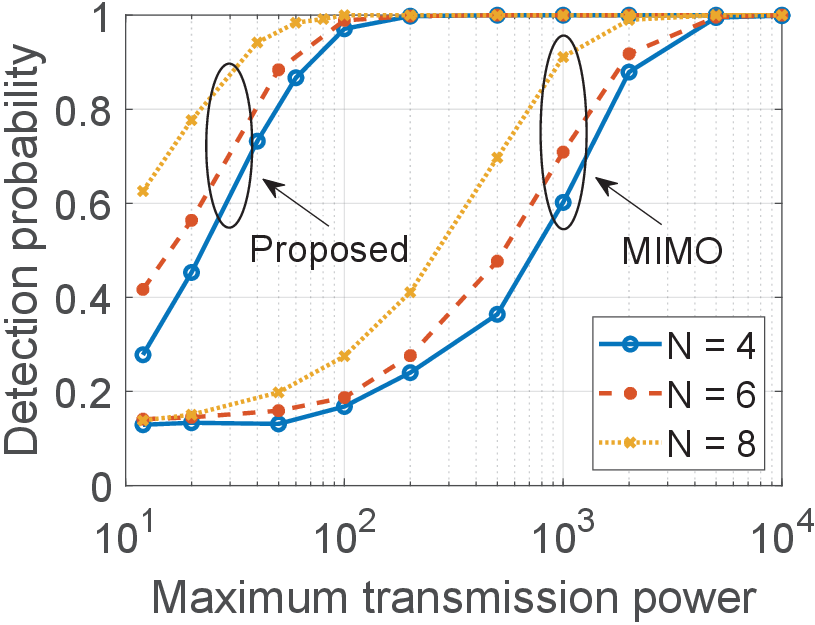}
  }
  \caption{
  (a) The detection probability $p(U_{j^*}|U_{j^*})$ versus the number of cycles $C$ in the detection process; (b) The mis-detection probability $\sum_{j\ne j^*}p(U_{j^*}|U_{j})p^1(U_j)$ versus the number of cycles $C$ in the detection process; (c) The detection probability $p(U_{j^*}|U_{j^*})$ versus the maximum transmission power $P_M$ with different numbers of antennas $N$.}
  \label{f_cp}
\end{figure*}

Fig.~\ref{f_cp} (a) and (b) show the detection probability $p(U_{j^*}|U_{j^*})$ and the mis-detection probability $\sum_{j\ne j^*}p(U_{j^*}|U_{j})p^1(U_j)$ versus the number of cycles $C$ in the detection process, respectively. It can be observed that the detection probability obtained by the proposed scheme is higher than those obtained by the random and MIMO schemes, and the mis-detection probability obtained by the proposed scheme is smaller than those obtained by the other two schemes, which verify the effectiveness of the proposed scheme. Besides, we can also observe that the performances of the random and proposed schemes are much higher than that of the MIMO scheme. Specifically, the detection probability obtained by the MIMO scheme increase slowly with the number of cycles and is lower than $0.2$ when $C = 20$, while the probabilities of detection obtained by the proposed and random schemes with RIS can approach $1$ after sufficient number of cycles. As for mis-detection, the probability obtained by the MIMO scheme when $C = 20$ is also much higher than those obtained by the random and proposed schemes. This is because by incorporating the RIS, the power gain of the radar can be significantly promoted even the phase shifts of the RIS are not optimized, leading to a rapid increase/decrease of detection/mis-detection probabilities of the random and proposed scheme.

Fig.~\ref{f_cp} (c) presents the detection probability $p(U_{j^*}|U_{j^*})$ versus the maximum transmission power $P_M$ when $C = 3$. It can be seen that for both the proposed and the MIMO schemes, the detection probability first increases and then remains close to $1$ when the maximum transmission power $P_M$ increases. Besides, the detection probability $p$ also increases with the number of antennas $N$. However, to reach the same detection probability, the maximum transmission power $P_M$ and the number of antennas $N$ required by the proposed scheme are much smaller than that of the MIMO scheme, which indicates a tradeoff between the cost of deploying an RIS and the cost of increasing the size and the power of MIMO antenna array.

\subsection{Radar Configuration}

\begin{figure*}[!t]
  \centering
  \subfloat[]{
    \includegraphics[height=1.8in]{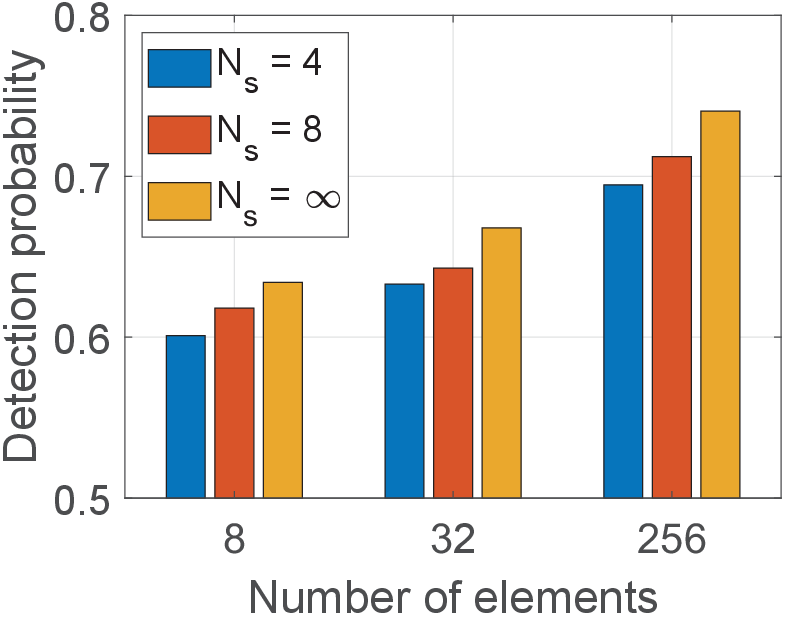}
  }
  \hspace{1in}
  \subfloat[]{
    \includegraphics[height=1.8in]{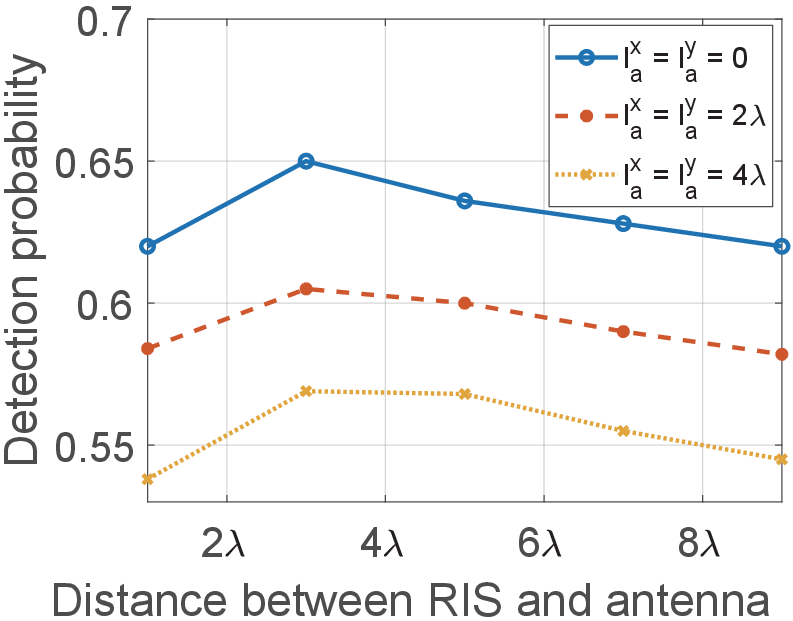}
  }
  \caption{(a) The detection probability $p(U_{j^*}|U_{j^*})$ versus the number of elements $M$ with different number of phase shifts $N_s$; (b) The detection probability $p(U_{j^*}|U_{j^*})$ versus the distance between the RIS and the antenna array $l^z_a$ with different $x$ and $y$ coordinates of the antenna array center.}
  \label{f_cp2}
\end{figure*}

Fig.~\ref{f_cp2} (a) shows the detection probability $p(U_{j^*}|U_{j^*})$ versus the number of elements $M$ with different number of phase shifts $N_s$ when $C = 6$. We can observe that the detection probability increases with the number of elements, which is in accordance with the conclusion in Section~\ref{sss_configuration}. Besides, the detection probability also increases with the number of phase shifts $N_s$, since the reflection coefficients of the RIS with more number of phase shifts can be adjusted more precisely to obtain a larger relative entropy and higher accuracy. This also indicates a trade-off between the number of elements and the number of phase shifts given the detection performance.

Fig.~\ref{f_cp2} (b) depicts the detection probability $p(U_{j^*}|U_{j^*})$ versus the distance between the RIS and the antenna array $l^z_a$ with different $x$ and $y$ coordinates of the antenna array center when $C = 6$. It can be observe that the detection probability when the center of the antenna array is close to the $z$ axis ($l^x_a = l^y_a = 0$) is higher than those when the center of the antenna array is far away from the the $z$ axis ($l^x_a = l^y_a = 2\lambda, 4\lambda$), which is in accordance with \textbf{Proposition~\ref{pro_opt_l_x_a}}. Besides, we can observe that the detection probability first increases and then decreases when the distance between the RIS and the antenna array $l^z_a$ increases, and there exists an optimal distance between the RIS and the antenna array with the best detection performance.

\subsection{Complexity}

\begin{figure*}[!t]
  \centering
  \subfloat[]{
      \includegraphics[height=1.8in]{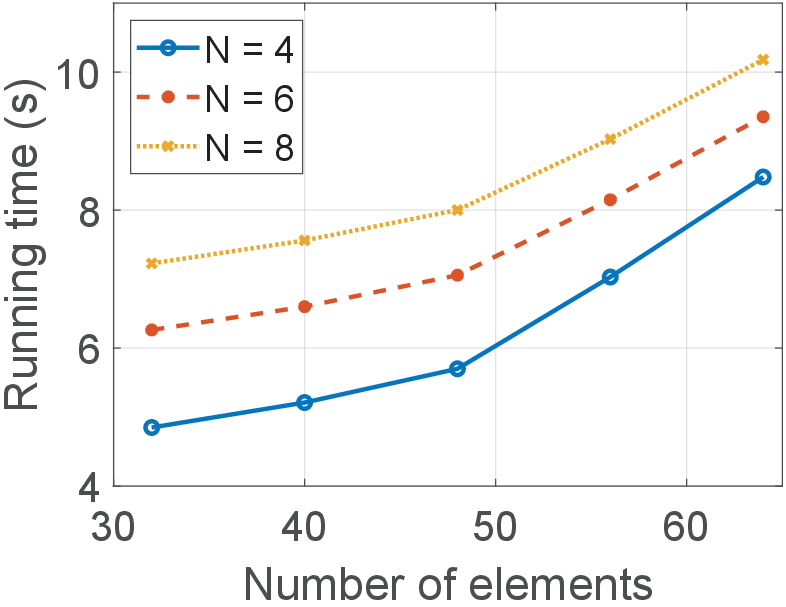}
  }
  \hspace{1in}
  \subfloat[]{
      \includegraphics[height=1.8in]{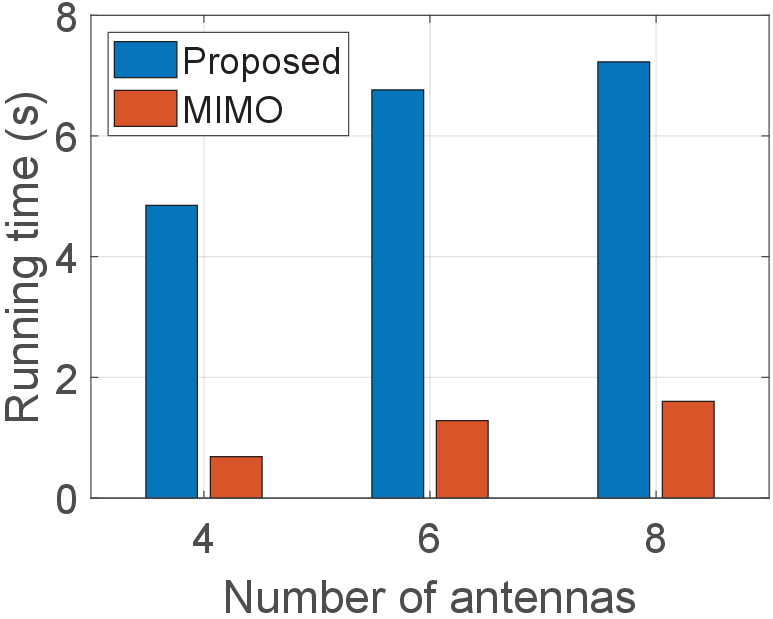}
  }
  \caption{(a) The running time $t^r$ versus the number of RIS elements $M$ with different numbers of antennas $N$; \textcolor{black}{(b) The running time $t^r$ versus the number of elements $N$ with different schemes.}}
  \label{f_cp3}
\end{figure*}

Fig.~\ref{f_cp3} (a) shows the running time $t^r$ of the WPSO algorithm in each cycle versus the number of RIS elements $M$, where the running time is obtained using a computer with Intel Core i5-8250U CPU (1.6GHz), 8 GB RAM, and Matlab 2019b.
We can observe that the running time increases with the number of RIS elements $M$ and the number of antennas $N$, which matches the results in Section~\ref{sss_complexity_analysis}.

\textcolor{black}{The running time $t^r$ in each cycle versus the number of elements $N$ with different schemes is shown in Fig.~\ref{f_cp3} (b). We can observe that the running time increases with the number of antennas for both schemes, and the running time of the proposed scheme is significantly longer than that of the traditional MIMO radar scheme. This is because the traditional MIMO radar scheme only optimizes the radar waveforms, while the proposed scheme iteratively optimizes the radar waveforms and the RIS phase shifts, which leads to a higher detection accuracy compared with the traditional MIMO radar scheme.}

\section{Conclusion}
\label{s_conclusion}

In this paper, we have investigated the multi-target detection using the RIS-assisted radar systems. A multi-target detection protocol has been designed to adaptively promote the detection performance by coordinating the operations of the antenna arrays and the RIS. We have also formulated the optimization problem for target detection based on the relative entropy criterion. To tackle the formulated problem, we have proposed the WPSO algorithm which jointly optimizes the radar waveforms and the RIS phase shifts. The detection performance of the proposed scheme has been analyzed theoretically, and the superiority of the proposed scheme has been verified through simulation. We can conclude from the results of analysis and the simulation that: 1) by incorporating an RIS in the radar systems which creates reflection paths between the antennas and the targets, the power gain of the radar can be enhanced, leading to a significant performance improvement of the MetaRadar compared with traditional radar schemes; 2) to meet the same detection performance, we can increase the number of RIS elements to reduce the demand for the number of phase shifts, and vice versa; 3) to maximize the detection accuracy, it is preferred to put the RIS and the antenna array coaxially, and the distance between the RIS and the antenna array needs to be carefully chosen.

\appendices

\section{Proof of Proposition \ref{pro_d}}
\label{proof_d}

Based on (\ref{def_p_y_c}) and (\ref{def_D}),
the predicted distance between hypotheses $U_j$ and $U_{j'}$ can be expressed as\small
\begin{align}
  &d(U_j, U_{j'}|\mathcal{P}) \notag\\
  =& \int p(\bm{y}|U_j, \mathcal{P}) \log \dfrac{p(\bm{y}|U_j, \mathcal{P})}{p(\bm{y}|U_{j'}, \mathcal{P})} d\bm{y} + \int p(\bm{y}|U_{j'}, \mathcal{P}) \log \dfrac{p(\bm{y}|U_{j'}, \mathcal{P})}{p(\bm{y}|U_{j}, \mathcal{P})} d\bm{y}\notag\\
  =& \dfrac{1}{2\sigma^2} \int p(\bm{y}|U_j, \mathcal{P}) \left(||\bm{y} - \overline{\bm{y}}(U_{j'}, \mathcal{P})||^2 - ||\bm{y} - \overline{\bm{y}}(U_{j}, \mathcal{P})||^2\right) d\bm{y}\notag\\
  &+ \dfrac{1}{2\sigma^2} \int p(\bm{y}|U_{j'}, \mathcal{P}) \left(||\bm{y} - \overline{\bm{y}}(U_{j}, \mathcal{P})||^2 - ||\bm{y} - \overline{\bm{y}}(U_{j'}, \mathcal{P})||^2\right) d\bm{y}\notag\\
  =& \dfrac{1}{\sigma^2} ||\overline{\bm{y}}(U_{j}, \mathcal{P}) - \overline{\bm{y}}(U_{j'}, \mathcal{P})||^2.\label{app_A_1}
\end{align}\normalsize

According to (\ref{def_y_c}), $\overline{\bm{y}}(U_{j}, \mathcal{P}) = \bm{F}(U_{j}, \mathcal{P})\hat{\bm{\gamma}}(U_{j})$, and thus we have
\begin{align}
  d(U_j, U_{j'}|\mathcal{P}) =& \dfrac{1}{\sigma^2} ||\bm{F}(U_{j}, \mathcal{P})\hat{\bm{\gamma}}(U_{j}) - \bm{F}(U_{j'}, \mathcal{P})\hat{\bm{\gamma}}(U_{j'})||^2\notag\\
  =& \dfrac{1}{\sigma^2} \text{Re}\Big(\hat{\bm{\gamma}}^{\text{H}}(U_{j})\bm{F}^{\text{H}}(U_{j}, \mathcal{P})\bm{F}(U_{j}, \mathcal{P})\hat{\bm{\gamma}}(U_{j}) \notag\\
  &+ \hat{\bm{\gamma}}^{\text{H}}(U_{j'})\bm{F}^{\text{H}}(U_{j'}, \mathcal{P})\bm{F}(U_{j'}, \mathcal{P})\hat{\bm{\gamma}}(U_{j'}) \notag\\
  &- 2 \hat{\bm{\gamma}}^{\text{H}}(U_{j})\bm{F}^{\text{H}}(U_j, \mathcal{P})\bm{F}(U_{j'}, \mathcal{P})\hat{\bm{\gamma}}(U_{j'})\Big) \label{app_A_2}
\end{align}

The term $\hat{\bm{\gamma}}^{\text{H}}(U_{j})\bm{F}^{\text{H}}(U_j, \mathcal{P})\bm{F}(U_{j'}, \mathcal{P})\hat{\bm{\gamma}}(U_{j'})$ in (\ref{app_A_2}) can be expressed as
\begin{align}
  &\hat{\bm{\gamma}}^{\text{H}}(U_{j})\bm{F}^{\text{H}}(U_j, \mathcal{P})\bm{F}(U_{j'}, \mathcal{P})\hat{\bm{\gamma}}(U_{j'}) \notag\\
  =& \text{tr}\Bigg(\bm{w}\bm{w}^{H}\sum^{N(U_j)}_{i=1}\sum^{N(U_{j'})}_{i'=1} \hat{\gamma}^{\text{H}}_i(U_j)\hat{\gamma}_{i'}(U_{j'})\bm{Q}^{\text{H}}_i(U_j, \bm{s}^t, \bm{s}^r) \notag\\
  &\times \bm{Q}_{i'}(U_{j'}, \bm{s}^t, \bm{s}^r) \Bigg)\notag\\
  =&\text{tr}\big(\bm{X}_{w}\bm{Z}''(U_j, U_{j'}, \bm{s}^t, \bm{s}^r)\big),
\end{align}
where $\bm{X}_w \!= \bm{w}\bm{w}^{H}$ and $\bm{Z}''(U_j, U_{j'}, \bm{s}^t, \bm{s}^r) \! =\! \sum^{N(U_j)}_{i=1} \!\sum^{N(U_{j'})}_{i'=1}\! \hat{\gamma}^{\text{H}}_i(U_j)\hat{\gamma}_{i'}(U_{j'})\bm{Q}^{\text{H}}_i(U_j, \bm{s}^t, \bm{s}^r) \bm{Q}_{i'}(U_{j'}, \bm{s}^t, \bm{s}^r)$. 

Therefore, the predicted distance between hypotheses $U_j$ and $U_{j'}$ is given by
\begin{align}
  d(U_j, U_{j'}|\mathcal{P}) =& \dfrac{1}{\sigma^2} \text{Re}\Big(\text{tr}\big(\bm{X}_w \bm{Z}''(U_j, U_{j}, \bm{s}^t, \bm{s}^r) \notag\\
  &+ \bm{X}_w \bm{Z}''(U_{j'}, U_{j'}, \bm{s}^t, \bm{s}^r) \notag\\
  &- 2\bm{X}_{w}\bm{Z}''(U_j, U_{j'}, \bm{s}^t, \bm{s}^r)\big)\Big)\notag\\
  =& \dfrac{1}{\sigma^2}\text{Re}\Big(\text{tr}(\bm{X}_w \bm{Z}'(U_j, U_{j'}, \bm{s}^t, \bm{s}^r))\Big)\notag\\
  =& \dfrac{1}{\sigma^2}\text{Re}\Big(\text{tr}(\bm{w}^{\text{H}} \bm{Z}'(U_j, U_{j'}, \bm{s}^t, \bm{s}^r)\bm{w})\Big),
\end{align}
where $\bm{Z}'(U_j, U_{j'}, \bm{s}^t, \bm{s}^r) = \bm{Z}''(U_j, U_{j}, \bm{s}^t, \bm{s}^r) + \bm{Z}''(U_{j'}, U_{j'}, \bm{s}^t, \bm{s}^r) - 2\bm{Z}''(U_j, U_{j'}, \bm{s}^t, \bm{s}^r)$.

\section{Proof of Proposition \ref{pro_d2}}
\label{proof_d2}

Based on (\ref{def_b}), (\ref{def_Y_c}), and (\ref{def_y_c}), the parameter $\bm{F}$ can be expressed as
\begin{align}
  \bm{F}(U_j, \mathcal{P}) =& (\bm{Q}'_1(U_j, \bm{w}, \bm{s}^r) \bm{r}'(\bm{s}^t) + \bm{\zeta}_1(U_j), \cdots, \notag\\
  &\bm{Q}'_K(U_j, \bm{w}, \bm{s}^r) \bm{r}'(\bm{s}^t) + \bm{\zeta}_K(U_j))\in \mathbb{C}^{NL_R \times K}, 
\end{align}
where $\bm{Q}'_k(U_j, \bm{w}, \bm{s}^r) = (\bm{J}^{\text{T}}_k \bm{W}^{\text{T}} \bm{H}^{\text{T}}) \otimes ((\bm{b}^{\text{T}}_k(\bm{s}^r) + \bm{\xi}^{\text{T}}_k)\bm{a}_k) \in \mathbb{C}^{NL_R\times M^2}$, $\bm{r}'(\bm{s}^t) = \text{vec}(\bm{R}(\bm{s}^t)) \in \mathbb{C}^{M^2\times 1}$, $\bm{\zeta}_k(U_j) = \text{vec}((\bm{b}^{\text{T}}_k(\bm{s}^r) + \bm{\xi}^{\text{T}}_k) \bm{\xi}_k \bm{W}^c \bm{J}_k)$, and $K = N(U_j)$. Since $\bm{R}(\bm{s}^t)$ is a diagonal matrix with diagonal elements being $\bm{r}(\bm{s}^t) = (r(s_1), \cdots, r(s_M))$, the $i$-th element of vector $\bm{r}'(\bm{s}^t)$ is given by
\begin{equation}
  \bm{r}'(\bm{s}^t)_i =
  \begin{cases}
    r_m(s_m), &i = (m-1)(M+1) + 1 \\
    0, &\text{otherwise.}
  \end{cases}
\end{equation}
Thus, we have
\begin{equation}
  \bm{Q}'_k(U_j, \bm{w}, \bm{s}^r) \bm{r}'(\bm{s}^t) = \bm{Q}_k(U_j, \bm{w}, \bm{s}^r) \bm{r}(\bm{s}^t),
\end{equation}
where $\bm{Q}_k(U_j, \bm{w}, \bm{s}^r)$ is given by
\begin{align}
  \bm{Q}_k(U_j, \bm{w}, \bm{s}^r) =& (\bm{Q}'_{k, 1}(U_j, \bm{w}, \bm{s}^r), \bm{Q}'_{k, M+2}(U_j, \bm{w}, \bm{s}^r), \cdots, \notag\\
  &\bm{Q}'_{k, M^2}(U_j, \bm{w}, \bm{s}^r)),
\end{align}
where $\bm{Q}'_{k, i}(U_j, \bm{w}, \bm{s}^r)$ denote the $i$-th column of $\bm{Q}'_k(U_j, \bm{w}, \bm{s}^r)$. Consequently, parameter $\bm{F}$ can be expressed as
\begin{align}
  \bm{F}(U_j, \mathcal{P}) =& (\bm{Q}_1(U_j, \bm{w}, \bm{s}^r) \bm{r}(\bm{s}^t) + \bm{\zeta}_1(U_j), \cdots, \notag\\
  &\bm{Q}_K(U_j, \bm{w}, \bm{s}^r) \bm{r}(\bm{s}^t) + \bm{\zeta}_K(U_j)).
\end{align}

Similar to the proof in Appendix~\ref{proof_d}, 
(\ref{app_A_2}), the predicted distance between hypotheses $U_j$ and $U_{j'}$ can be expressed as
\begin{align}
  d(U_j, U_{j'}|\mathcal{P}) =& \dfrac{1}{\sigma^2} \text{Re}\Big(\hat{\bm{\gamma}}^{\text{H}}(U_{j})\bm{F}^{\text{H}}(U_{j}, \mathcal{P})\bm{F}(U_{j}, \mathcal{P})\hat{\bm{\gamma}}(U_{j}) \notag\\
  &+ \hat{\bm{\gamma}}^{\text{H}}(U_{j'})\bm{F}^{\text{H}}(U_{j'}, \mathcal{P})\bm{F}(U_{j'}, \mathcal{P})\hat{\bm{\gamma}}(U_{j'}) \notag\\
  &- 2 \hat{\bm{\gamma}}^{\text{H}}(U_{j})\bm{F}^{\text{H}}(U_j, \mathcal{P})\bm{F}(U_{j'}, \mathcal{P})\hat{\bm{\gamma}}(U_{j'})\Big),
\end{align}
where $\hat{\bm{\gamma}}^{\text{H}}(U_{j})\bm{F}^{\text{H}}(U_j, \mathcal{P})\bm{F}(U_{j'}, \mathcal{P})\hat{\bm{\gamma}}(U_{j'})$ can be expressed as
\begin{align}
  &\hat{\bm{\gamma}}^{\text{H}}(U_{j})\bm{F}^{\text{H}}(U_j, \mathcal{P})\bm{F}(U_{j'}, \mathcal{P})\hat{\bm{\gamma}}(U_{j'}) \notag\\
  =~& \sum^{N(U_j)}_{i=1} \sum^{N(U_j')}_{i'=1} \hat{\gamma}^{\text{H}}_i(U_j)\hat{\gamma}_{i'}(U_{j'})\big(\bm{r}^{\text{H}}(\bm{s}^t) \bm{Q}^{\text{H}}_i(U_j, \bm{w}, \bm{s}^r) \notag\\
  &+ \bm{\zeta}^{\text{H}}_i(U_j)\big) \left(\bm{Q}_{i'}(U_{j'}, \bm{w}, \bm{s}^r) \bm{r}(\bm{s}^t) + \bm{\zeta}_{i'}(U_{j'})\right) \notag\\
  =~& \bm{r}^{\text{H}} \bm{Z}''(U_j, U_{j'}, \bm{w}, \bm{s}^r) \bm{r}(\bm{s}^t) + \bm{r}^{\text{H}} \bm{z}''_1(U_j, U_{j'}, \bm{w}, \bm{s}^r) \notag\\
  &+ \bm{z}''_2(U_j, U_{j'}, \bm{w}, \bm{s}^r) \bm{r}(\bm{s}^t) + z''_3(U_j, U_{j'}),\notag
\end{align}
where 
\begin{align}
\bm{Z}''(U_j, U_{j'}, \bm{w}, \bm{s}^r) =& \sum^{N(U_j)}_{i=1}\sum^{N(U_{j'})}_{i'=1} \hat{\gamma}^{\text{H}}_i(U_j)\hat{\gamma}_{i'}(U_{j'})\notag\\
&\bm{Q}^{\text{H}}_i(U_j, \bm{w}, \bm{s}^r) \bm{Q}_{i'}(U_{j'}, \bm{w}, \bm{s}^r),\\
\bm{z}''_1(U_j, U_{j'}, \bm{w}, \bm{s}^r) =& \sum^{N(U_j)}_{i=1}\sum^{N(U_{j'})}_{i'=1} \hat{\gamma}^{\text{H}}_i(U_j)\hat{\gamma}_{i'}(U_{j'})\notag\\
&\bm{Q}^{\text{H}}_i(U_j, \bm{w}, \bm{s}^r) \bm{\zeta}_{i'}(U_j'),\\
\bm{z}''_2(U_j, U_{j'}, \bm{w}, \bm{s}^r) =& \sum^{N(U_j)}_{i=1}\sum^{N(U_{j'})}_{i'=1} \hat{\gamma}^{\text{H}}_i(U_j)\hat{\gamma}_{i'}(U_{j'})\bm{\zeta}^{\text{H}}_i(U_j) \notag\\
&\bm{Q}_{i'}(U_{j'}, \bm{w}, \bm{s}^r),\\
z''_3(U_j, U_{j'}) =& \sum^{N(U_j)}_{i=1}\sum^{N(U_{j'})}_{i'=1}\hat{\gamma}^{\text{H}}_i(U_j)\hat{\gamma}_{i'}(U_{j'})\notag\\
&\bm{\zeta}^{\text{H}}_i(U_j) \bm{\zeta}_{i'}(U_{j'}).
\end{align}

Therefore, the predicted distance between hypotheses $U_j$ and $U_{j'}$ is given by
\begin{align}
  d(U_j, U_{j'}|\mathcal{P}) =& \dfrac{1}{\sigma^2} \text{Re}\Big(\hat{\bm{\gamma}}^{\text{H}}(U_{j})\bm{F}^{\text{H}}(U_{j}, \mathcal{P})\bm{F}(U_{j}, \mathcal{P})\hat{\bm{\gamma}}(U_{j}) \notag\\
  &+ \hat{\bm{\gamma}}^{\text{H}}(U_{j'})\bm{F}^{\text{H}}(U_{j'}, \mathcal{P})\bm{F}(U_{j'}, \mathcal{P})\hat{\bm{\gamma}}(U_{j'}) \notag\\
  &- 2 \hat{\bm{\gamma}}^{\text{H}}(U_{j})\bm{F}^{\text{H}}(U_j, \mathcal{P})\bm{F}(U_{j'}, \mathcal{P})\hat{\bm{\gamma}}(U_{j'})\Big)\notag\\
  =& \dfrac{1}{\sigma^2}\text{Re}\Big(\bm{r}^{\text{H}}(\bm{s}^t) \bm{Z}'(U_j, U_{j'}, \bm{w}, \bm{s}^r)\bm{r}(\bm{s}^t) \notag\\
  &+ \bm{r}^{\text{H}}(\bm{s}^t) \bm{z}'_1(U_j, U_{j'}, \bm{w}, \bm{s}^r) \notag\\
  &+ \bm{z}'_2(U_j, U_{j'}, \bm{w}, \bm{s}^r) \bm{r}(\bm{s}^t) + z'_3(U_j, U_{j'}) \Big),
\end{align}
where 
\begin{align}
\bm{Z}'(U_j, U_{j'}, \bm{w}, \bm{s}^r) =& \bm{Z}''(U_j, U_{j}, \bm{w}, \bm{s}^r) + \bm{Z}''(U_{j'}, U_{j'}, \bm{w}, \bm{s}^r) \notag\\
&- 2\bm{Z}''(U_j, U_{j'}, \bm{w}, \bm{s}^r),\\
\bm{z}'_1(U_j, U_{j'}, \bm{w}, \bm{s}^r) =&  \bm{z}''_1(U_j, U_{j}, \bm{w}, \bm{s}^r) + \bm{z}''_1(U_{j'}, U_{j'}, \bm{w}, \bm{s}^r) \notag\\
&- 2\bm{z}''_1(U_j, U_{j'}, \bm{w}, \bm{s}^r),\\
\bm{z}'_2(U_j, U_{j'}, \bm{w}, \bm{s}^r) =& \bm{z}''_2(U_j, U_{j}, \bm{w}, \bm{s}^r) + \bm{z}''_2(U_{j'}, U_{j'}, \bm{w}, \bm{s}^r) \notag\\
&- 2\bm{z}''_2(U_j, U_{j'}, \bm{w}, \bm{s}^r),\\
z'_3 (U_j, U_{j'}) =& z''_3(U_j, U_{j}) + z''_3(U_{j'}, U_{j'}) - 2z''_3(U_j, U_{j'}).
\end{align}

\section{Proof of Proposition \ref{pro_obj_upper}}
\label{proof_obj_upper}

Based on~(\ref{def_beta}) and (\ref{app_A_1}), the objective function~(\ref{p1_obj}) can be expressed as
\begin{align}
  &\sum^{J-1}_{j = 0}\sum^{J-1}_{j' = j+1}\beta_{j, j'} d(U_{j}, U_{j'}|\mathcal{P}) \notag\\
  =& \sum^{J-1}_{j = 0}\sum^{J-1}_{j' = j+1} \dfrac{\beta_{j, j'}}{\sigma^2} ||\overline{\bm{y}}(U_{j}, \mathcal{P}) - \overline{\bm{y}}(U_{j'}, \mathcal{P})||^2\notag\\
  \le& \sum^{J-1}_{j = 0}\sum^{J-1}_{j' = j+1} \dfrac{\beta_{j, j'}}{\sigma^2} \left(||\overline{\bm{y}}(U_{j}, \mathcal{P})||^2 + ||\overline{\bm{y}}(U_{j'}, \mathcal{P})||^2\right),\notag\\
  \le& \sum^{J-1}_{j = 0}\sum^{J-1}_{j' = j+1} \dfrac{1}{\sigma^2} \left(||\overline{\bm{y}}(U_{j}, \mathcal{P})||^2 + ||\overline{\bm{y}}(U_{j'}, \mathcal{P})||^2\right).
\end{align}

Therefore, the objective function~(\ref{p1_obj}) has an upper bound if the energy of the received signals $||\overline{\bm{y}}(U_{j}, \mathcal{P})||^2$ is limited. Since the energy of signals will decrease due to the path loss and the reflection by the RIS and the targets, the energy of the received signals cannot be greater than the that of the transmitted signals $P_M$. Therefore, we have
\begin{align}
  \sum^{J-1}_{j = 0}\sum^{J-1}_{j' = j+1}\beta_{j, j'} d(U_{j}, U_{j'}|\mathcal{P}) \le& \sum^{J-1}_{j = 0}\sum^{J-1}_{j' = j+1} \dfrac{2P_M}{\sigma^2} = \dfrac{J(J-1)P_M}{2}.
\end{align}

\section{Proof of Proposition \ref{pro_opt_l_x_a}}
\label{proof_opt_l_x_a}

The power gain of the MetaRadar when the isotropic antenna is at the location $(l^x_a, l^z_a)$ is given by
\begin{align}
  &B(l^x_a, l^z_a)\notag\\
  =& \left(\sum^M_{m=1} \dfrac{\rho G^R_P(\theta^r_{1, m})}{l_{1, m}} + 1\right)^4\notag\\
  =& \left(\sum^M_{m=1} \dfrac{\rho\text{cos}^{1.5}(\theta^r_{1, m})}{l_{1, m}} + 1\right)^4\notag\\
  =& \left(\sum^M_{m=1} \dfrac{\rho (l^z_a)^{1.5}}{((l^z_a)^2 + (l^x_a - l^x_m)^2)^{1.25}} + 1\right)^4\notag\\
  =& \left(\sum^M_{m=1} \dfrac{(\rho l^z_a)^{1.5}}{((l^z_a)^2 + (l^x_a + (M+1)l^e/2 - m l^e)^2)^{1.25}} + 1\right)^4
\end{align}

Suppose $l^x_a = n l^e + l'$, where $n$ is an non-zero integer and $l' \in [-l^e/2, l^e/2]$. It can be proved that $B(l', l^z_a) > B(l^x_a, l^z_a)$ always holds. Specifically, the power gain when the antenna is at $(l^x_a, l^z_a)$ can be expressed as\small
\begin{align}
  &B(l^x_a, l^z_a)\notag\\
  =& \left(\sum^M_{m=1} \dfrac{\rho(l^z_a)^{1.5}}{((l^z_a)^2 + (l' + (M+1)l^e/2 - (m-n) l^e)^2)^{1.25}} + 1\right)^4,
\end{align}\normalsize
and the power gain when the antenna is at $(l', l^z_a)$ can be expressed as\small
\begin{align}
  B(l', l^z_a) =& \left(\sum^M_{m=1} \dfrac{\rho (l^z_a)^{1.5}}{((l^z_a)^2 + (l' + (M+1)l^e/2 - m l^e)^2)^{1.25}} + 1\right)^4\label{app_d_3}.
\end{align}\normalsize

To prove $B(l^x_a, l^z_a) > B(l^x_a, l^z_a)$, it is equivalent to prove the following relationship hold.
\begin{align}
  &\sum^M_{m=1} \dfrac{(l^z_a)^{1.5}}{((l^z_a)^2 \!+\! (l' \!+\! (M\!+\!1)l^e/2 \!-\! m l^e)^2)^{1.25}} \notag\\
  >\!& \sum^M_{m=1} \dfrac{(l^z_a)^{1.5}}{((l^z_a)^2 \!+\! (l' \!+\! (M\!+\!1)l^e/2 \!-\! (m\!-\!n) l^e)^2)^{1.25}}\label{app_d_4}.
\end{align}
When $M = 1$, it is obvious that (\ref{app_d_4}) holds. When $n \ge \lfloor M/2 \rfloor$ and $M > 1$, it is also easy to prove (\ref{app_d_4}). When $0<n< \lfloor M/2 \rfloor$ and $M > 1$, we have\small
\begin{align}
  &\sum^M_{m=1} \dfrac{(l^z_a)^{1.5}}{((l^z_a)^2 + (l' + (M+1)l^e/2 - m l^e)^2)^{1.25}} \notag\\
  &- \sum^M_{m=1} \dfrac{(l^z_a)^{1.5}}{((l^z_a)^2 + (l' + (M+1)l^e/2 - (m-n) l^e)^2)^{1.25}}\notag\\
  =& \sum^{M-n}_{m=1} \dfrac{(l^z_a)^{1.5}}{((l^z_a)^2 + (l' + (M+1)l^e/2 - m l^e)^2)^{1.25}} \notag\\
  &- \sum^M_{m=n+1} \dfrac{(l^z_a)^{1.5}}{((l^z_a)^2 + (l' + (M+1)l^e/2 - (m-n) l^e)^2)^{1.25}}\notag\\
  & + \sum^{M}_{m=M-n+1} \dfrac{(l^z_a)^{1.5}}{((l^z_a)^2 + (l' + (M+1)l^e/2 - m l^e)^2)^{1.25}} \notag\\
  &- \sum^{n}_{m=1} \dfrac{(l^z_a)^{1.5}}{((l^z_a)^2 + (l' + (M+1)l^e/2 - (m-n) l^e)^2)^{1.25}}\notag\\
  =& \sum^{n}_{m=1} \dfrac{(l^z_a)^{1.5}}{((l^z_a)^2 + (l' + (M+1)l^e/2 - (m - n + M) l^e)^2)^{1.25}} \notag\\
  &- \dfrac{(l^z_a)^{1.5}}{((l^z_a)^2 + (l' + (M+1)l^e/2 - (m-n) l^e)^2)^{1.25}}.\label{app_d_5}
\end{align}\normalsize
Since $0<n< \lfloor M/2 \rfloor$, we have\small
\begin{align}
  &|(l^z_a)^2 + (l' + (M+1)l^e/2 - (m - n + M) l^e)^2| \notag\\
  <& |(l^z_a)^2 + (l' + (M+1)l^e/2 + (n-m ) l^e)^2|,
\end{align}\normalsize
and thus (\ref{app_d_5}) is greater than $0$. Similar process can be applied to prove the case when $n < 0$, which is omitted here for brevity. Since $B(l^x_a, l^z_a) > B(l^x_a, l^z_a)$ holds, the $l' \in [-l^e/2, l^e/2]$ with higher power gain than that of $l^x_a$ has been found.


%








\begin{IEEEbiography}[{\includegraphics[width=1in,height=1.25in,clip,keepaspectratio]{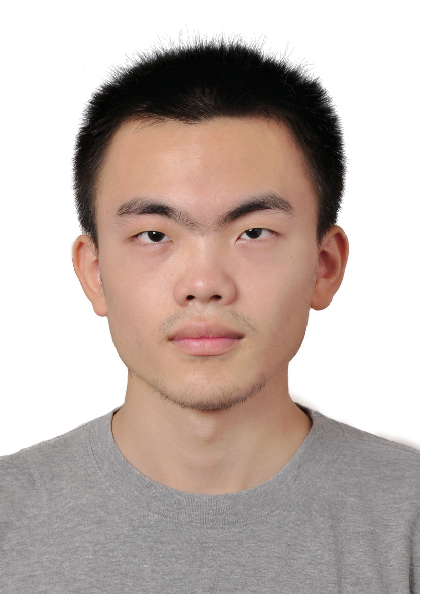}}]{Haobo Zhang} (S’19) received the B.S. degree at School of Electrical Engineering and Computer Science in Peking University in 2019, where he is currently pursuing the PhD degree in signal and information processing. His research interests include metasurface, wireless networks, and optimization theory.
\end{IEEEbiography}

\begin{IEEEbiography}[{\includegraphics[width=1in,height=1.25in,clip,keepaspectratio]{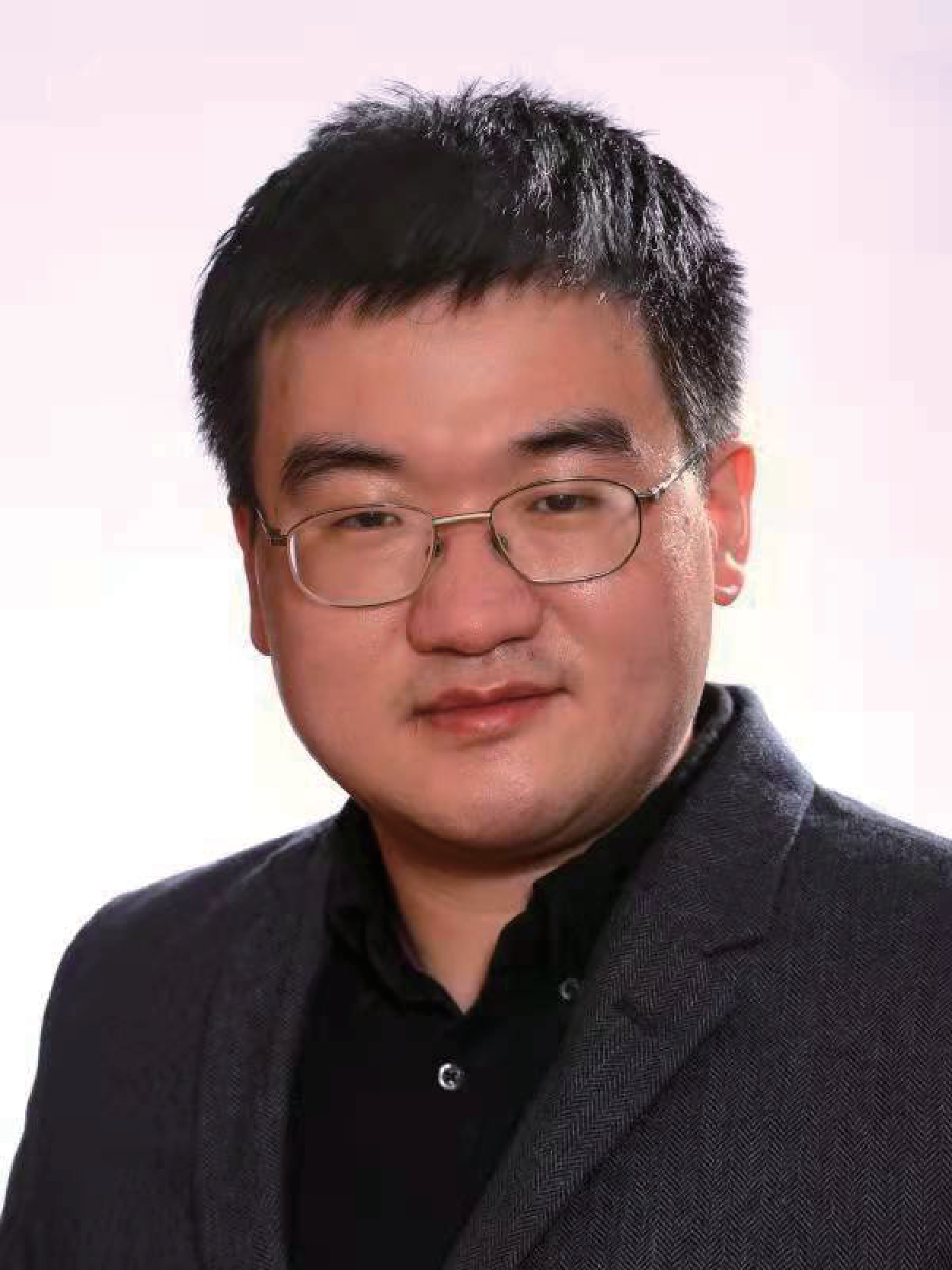}}]{Hongliang Zhang} (S’15-M’19) received the B.S. and Ph.D. degrees at the School of Electrical Engineering and Computer Science at Peking University, in 2014 and 2019, respectively. He was a Postdoctoral Fellow in the Electrical and Computer Engineering Department at the University of Houston, Texas. Currently, he is a Postdoctoral Associate in the Department of Electrical and Computer Engineering at Princeton University, New Jersey. His current research interest includes reconfigurable intelligent surfaces, aerial access networks, optimization theory, and game theory. He received the best doctoral thesis award from Chinese Institute of Electronics in 2019. He is an exemplary reviewer for IEEE Transactions on Communications in 2020. He is also the recipient of 2021 IEEE Comsoc Heinrich Hertz Award for Best Communications Letters and 2021 IEEE ComSoc Asia-Pacific Outstanding Paper Award. He has served as a TPC Member for many IEEE conferences, such as Globecom, ICC, and WCNC. He is currently an Editor for IEEE Communications Letters, IET Communications, and Frontiers in Signal Processing. He has also served as a Guest Editor for several journals, such as IEEE Internet of Things Journal, Journal of Communications and Networks, etc.
\end{IEEEbiography}

\begin{IEEEbiography}[{\includegraphics[width=1in,height=1.25in,clip,keepaspectratio]{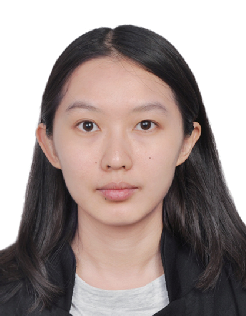}}]{Boya Di} (S’15-M’19) obtained her PhD degree from the Department of Electronics, Peking University, China, in 2019. Prior to that, she received the B.S. degree in electronic engineering from Peking University in 2014. She was a postdoc researcher at Imperial College London and is now an assistant professor at Peking University. Her current research interests include reconfigurable intelligent surfaces, multi-agent systems, edge computing, vehicular networks, and aerial access networks. She received the best doctoral thesis award from China Education Society of Electronics in 2019.  She is also the recipient of 2021 IEEE ComSoc Asia-Pacific Outstanding Paper Award. She serves as an associate editor for IEEE Transactions on Vehicular Technology since June 2020. She has also served as a workshop co-chair for IEEE WCNC 2020\&2021.
\end{IEEEbiography}

\begin{IEEEbiography}[{\includegraphics[width=1in,height=1.25in,clip,keepaspectratio]{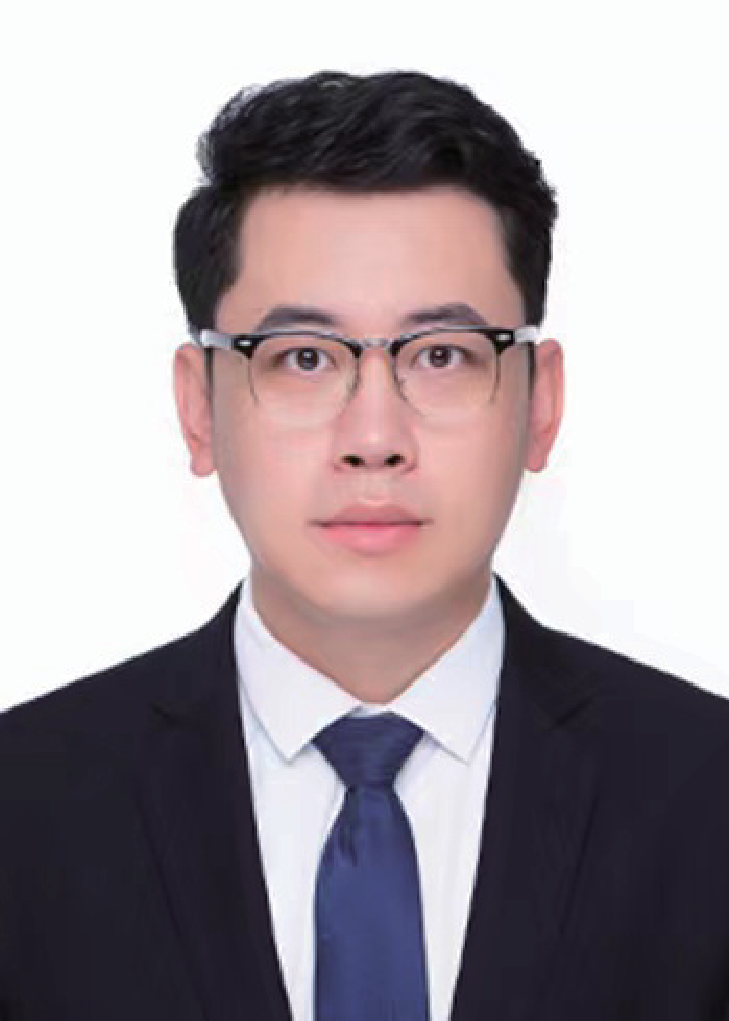}}]{Kaigui Bian} (S'05, M'11) received his Ph.D. degree in Computer Engineering from Virginia Tech in 2011, and his B.S. degree in Computer Science from Peking University, Beijing, China in 2005. He was a Visiting Young Faculty in Microsoft Research Asia in 2013. He received the best paper awards of international conferences (IEEE ICC 2015, ICCSE 2017, BIGCOM 2018) and the best student paper award of IEEE DSC 2018. He was the recipient of IEEE Communication Society Asia-Pacific Board (APB) Outstanding Young Researcher Award in 2018. He serves as an Editor for IEEE Transactions on Vehicular Technology, and the organizing committee member as well as technical program committee member of many international conferences. His research interests include wireless networking and mobile computing.
\end{IEEEbiography}

\begin{IEEEbiography}[{\includegraphics[width=1in,height=1.25in,clip,keepaspectratio]{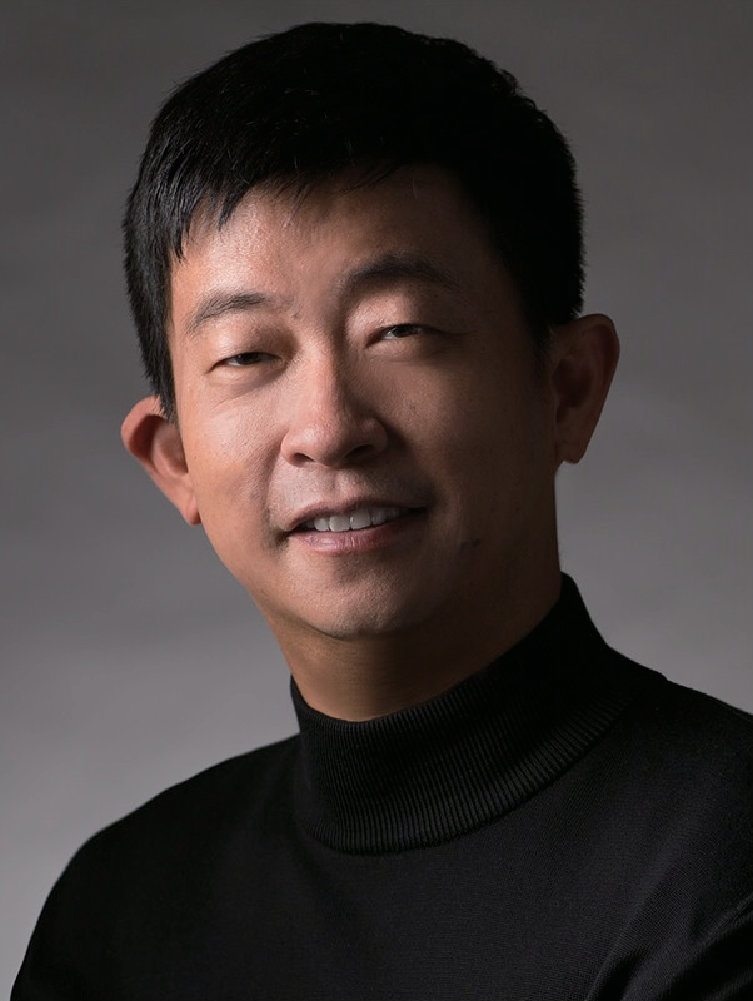}}]{Zhu Han}(S’01–M’04-SM’09-F’14) received the B.S. degree in electronic engineering from Tsinghua University, in 1997, and the M.S. and Ph.D. degrees in electrical and computer engineering from the University of Maryland, College Park, in 1999 and 2003, respectively. 

From 2000 to 2002, he was an R\&D Engineer of JDSU, Germantown, Maryland. From 2003 to 2006, he was a Research Associate at the University of Maryland. From 2006 to 2008, he was an assistant professor at Boise State University, Idaho. Currently, he is a John and Rebecca Moores Professor in the Electrical and Computer Engineering Department as well as in the Computer Science Department at the University of Houston, Texas. His research interests include wireless resource allocation and management, wireless communications and networking, game theory, big data analysis, security, and smart grid.  Dr. Han received an NSF Career Award in 2010, the Fred W. Ellersick Prize of the IEEE Communication Society in 2011, the EURASIP Best Paper Award for the Journal on Advances in Signal Processing in 2015, IEEE Leonard G. Abraham Prize in the field of Communications Systems (best paper award in IEEE JSAC) in 2016, and several best paper awards in IEEE conferences. Dr. Han was an IEEE Communications Society Distinguished Lecturer from 2015-2018, AAAS fellow since 2019, and ACM distinguished Member since 2019. Dr. Han is a 1\% highly cited researcher since 2017 according to Web of Science. Dr. Han is also the winner of the 2021 IEEE Kiyo Tomiyasu Award, for outstanding early to mid-career contributions to technologies holding the promise of innovative applications, with the following citation: ``for contributions to game theory and distributed management of autonomous communication networks."  
\end{IEEEbiography}

\begin{IEEEbiography}[{\includegraphics[width=1in,height=1.25in,clip,keepaspectratio]{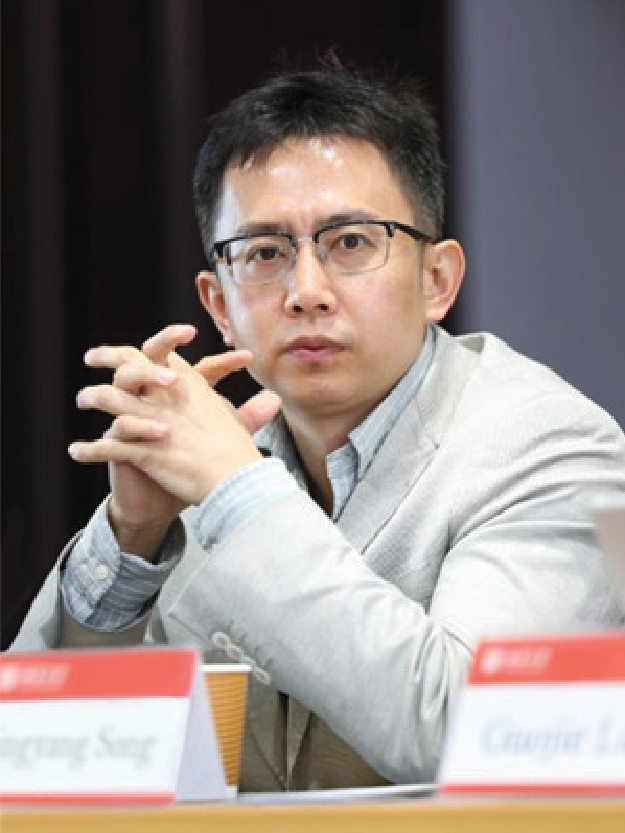}}]{Lingyang Song} (S’03–M’06-SM’11-F’19) received his PhD from the University of York, UK, in 2007, where he received the K. M. Stott Prize for excellent research. He worked as a postdoctoral research fellow at the University of Oslo, Norway, and Harvard University, until rejoining Philips Research UK in March 2008. In May 2009, he joined the School of Electronics Engineering and Computer Science, Peking University, China, as a full professor. His main research interests include cooperative and cognitive communications, physical layer security, and wireless ad hoc/sensor networks. He published extensively, wrote 6 text books, and is co-inventor of a number of patents (standard contributions). He received 9 paper awards in IEEE journal and conferences including IEEE JSAC 2016, IEEE WCNC 2012, ICC 2014, Globecom 2014, ICC 2015, etc. He is currently on the Editorial Board of IEEE Transactions on Wireless Communications and Journal of Network and Computer Applications. He served as the TPC co-chairs for the International Conference on Ubiquitous and Future Networks (ICUFN2011/2012), symposium co-chairs in the International Wireless Communications and Mobile Computing Conference (IWCMC 2009/2010), IEEE International Conference on Communication Technology (ICCT2011), and IEEE International Conference on Communications (ICC 2014, 2015). He is the recipient of 2012 IEEE Asia Pacific (AP) Young Researcher Award. Dr. Song is a fellow of IEEE, and IEEE ComSoc distinguished lecturer since 2015. 
\end{IEEEbiography}

\end{document}